\documentclass[a4paper,twocolumn,amsmath,amssymb,prb,preprintnumbers,notitlepage,preprintfigures,longbibliography]{revtex4-2}

\usepackage{graphicx}\usepackage{dcolumn}\usepackage{amsmath}
\usepackage{verbatim}
\usepackage[usenames,dvipsnames]{color}
\usepackage{bm}
\usepackage{bbm}
\usepackage{natbib}
\usepackage{xspace}
\usepackage{marginnote}
\usepackage{mathtools}
\usepackage{dsfont}
\usepackage{hyperref}
\usepackage[caption=false]{subfig}
\usepackage{tikz} \hypersetup{colorlinks=true,linktoc=all,linkcolor=blue,breaklinks=true,citecolor=blue,urlcolor=blue}
\usetikzlibrary{calc}
\usepackage{csquotes} 
\usepackage{etoolbox} \robustify{\uparrow}
\robustify{\downarrow}
\robustify{\sum}
\robustify{\int}
\robustify{\nonumber}
\robustify{\cite}
\robustify{\footnote}

\renewrobustcmd{\Re}{{\text{Re}}}
\renewrobustcmd{\Im}{{\text{Im}}}

\newrobustcmd{\sdag}{{\bm{\dag}}}  \newrobustcmd{\tot}{\text{tot}} \renewrobustcmd{\l}{\text{L}}   \newrobustcmd{\n}{\text{N}}   
\newrobustcmd{\deltah}{\bar{\delta}} \newrobustcmd{\sign}{\text{sgn}}

\newrobustcmd{\K}{\mathcal{K}}
\newrobustcmd{\G}{\mathcal{G}}
\newrobustcmd{\F}{\mathcal{F}} 
\newrobustcmd{\E}{\omega}
\renewrobustcmd{\S}{\mathcal{S}}

\renewrobustcmd{\H}{\mathcal{H}}   \newrobustcmd{\T}{\mathcal{T}} 
\newrobustcmd{\one}{\mathds{1}}   
\newrobustcmd{\ones}{\mathcal{I}}

\newrobustcmd{\ket}[1]{|#1\rangle}
\newrobustcmd{\bra}[1]{\langle#1|}
\newrobustcmd{\brkt}[1]{\langle #1 \rangle}
\newrobustcmd{\braket}[2]{\langle #1 | #2 \rangle}

\newrobustcmd{\Ket}[1]{\bm{|}#1\bm{)}}
\newrobustcmd{\Bra}[1]{\bm{(}#1\bm{|}}
\newrobustcmd{\Braket}[2]{\bm{(}#1\bm{|}#2\bm{)}}
\newrobustcmd{\Brkt}[1]{\bm{(} #1 \bm{)}}

\newrobustcmd{\op}[1]{\hat{#1}}
\newrobustcmd{\sop}[1]{\mathcal{#1}}

\DeclareMathOperator{\Tr}{Tr}
\newrobustcmd{\tr}[1]{\underset{#1}{\Tr}}
\newrobustcmd{\tri}[1]{{\Tr}_{#1}}

\newrobustcmd{\Krate}{\mathsf{k}}
\newrobustcmd{\Grate}{\mathsf{g}}
\newrobustcmd{\Prate}{\mathsf{p}}

\newrobustcmd{\Eq}[1]{Eq.~(\ref{#1})}
\newrobustcmd{\Eqs}[1]{Eqs.~(\ref{#1})}
\newrobustcmd{\eq}[1]{(\ref{#1})}
\newrobustcmd{\Fig}[1]{Fig.~\ref{#1}}
\newrobustcmd{\fig}[1]{\ref{#1}}
\newrobustcmd{\Figs}[1]{Figs.~\ref{#1}}
\newrobustcmd{\Sec}[1]{Sec.~\ref{#1}}
\newrobustcmd{\App}[1]{App.~\ref{#1}}
\newrobustcmd{\app}[1]{\ref{#1}}
\ifdefined\Ref
\renewrobustcmd{\Ref}[1]{Ref.~[\onlinecite{#1}]}
\else
\newrobustcmd{\Ref}[1]{Ref.~[\onlinecite{#1}]}
\fi
\newrobustcmd{\Refs}[1]{Refs.~[\onlinecite{#1}]}

\definecolor{green}{rgb}{0.0,0.75,0.0}
\definecolor{red}{rgb}{1.0,0.0,0.0}

\newcommand{\todo}[1]{}
\newcommand{\new}[1]{{#1}}

\usepackage{hyperref} \hypersetup{colorlinks=true,linktoc=all,linkcolor=blue,breaklinks=false,citecolor=blue,urlcolor=blue}

\usepackage{graphicx}

\begin{document}

\begin{tikzpicture}
\coordinate (lengthInnerBlock) at ($0.8*(1,0)$);
\coordinate (height) at ($0.8*(0,0.5)$);
\coordinate (smallHeight) at ($0.7*(height)$);
\coordinate (lengthOuterLeg) at (0.5,0);
\end{tikzpicture}

\newcommand{\DiagramKTwo}{
\begin{tikzpicture}[baseline={([yshift=-.5ex]current bounding box.center)}]
\coordinate (t) at (0,0);
\coordinate (t0) at (lengthInnerBlock);
\coordinate (tH) at ($(t)+(height)$);
\coordinate (t0H) at ($(t0)+(height)$);
\draw[thick] (t) -- (t0);
\draw[thick] (t) -- (tH) -- (t0H) -- (t0);
\filldraw [very thick] (t) circle (2pt);
\filldraw [very thick] (t0) circle (2pt);
\end{tikzpicture}
}

\newcommand{\DiagramKFourOne}{
\begin{tikzpicture}[baseline={([yshift=-.5ex]current bounding box.center)}]
\coordinate (t) at (0,0);
\coordinate (t0) at ($3*(lengthInnerBlock)$);
\coordinate (t1) at ($1*(lengthInnerBlock)$);
\coordinate (t2) at ($2*(lengthInnerBlock)$);
\coordinate (tH) at ($(t)+(height)$);
\coordinate (t0H) at ($(t0)+(smallHeight)$);
\coordinate (t1H) at ($(t1)+(smallHeight)$);
\coordinate (t2H) at ($(t2)+(height)$);
\draw[thick] (t) -- (t0);
\draw[thick] (t) -- (tH) -- (t2H) -- (t2);
\draw[thick] (t1) -- (t1H) -- (t0H) -- (t0);
\filldraw [very thick] (t) circle (2pt);
\filldraw [very thick] (t1) circle (2pt);
\filldraw [very thick] (t2) circle (2pt);
\filldraw [very thick] (t0) circle (2pt);
\end{tikzpicture}
}

\newcommand{\DiagramKFourTwo}{
\begin{tikzpicture}[baseline={([yshift=-.5ex]current bounding box.center)}]
\coordinate (t) at (0,0);
\coordinate (t0) at ($3*(lengthInnerBlock)$);
\coordinate (t1) at ($1*(lengthInnerBlock)$);
\coordinate (t2) at ($2*(lengthInnerBlock)$);
\coordinate (tH) at ($(t)+(height)$);
\coordinate (t0H) at ($(t0)+(height)$);
\coordinate (t1H) at ($(t1)+(smallHeight)$);
\coordinate (t2H) at ($(t2)+(smallHeight)$);
\draw[thick] (t) -- (t0);
\draw[thick] (t) -- (tH) -- (t0H) -- (t0);
\draw[thick] (t1) -- (t1H) -- (t2H) -- (t2);
\filldraw [very thick] (t) circle (2pt);
\filldraw [very thick] (t1) circle (2pt);
\filldraw [very thick] (t2) circle (2pt);
\filldraw [very thick] (t0) circle (2pt);
\end{tikzpicture}
}

\newcommand{\DiagramPiTwo}{
\begin{tikzpicture}[baseline={([yshift=-.5ex]current bounding box.center)}]
\coordinate (t) at (0,0);
\coordinate (t0) at ($(lengthInnerBlock) + 2*(lengthOuterLeg)$);
\coordinate (t1) at (lengthOuterLeg);
\coordinate (t2) at ($(lengthOuterLeg) + (lengthInnerBlock)$);
\coordinate (t1H) at ($(t1)+(height)$);
\coordinate (t2H) at ($(t2)+(height)$);
\draw[thick] (t) -- (t0);
\draw[thick] (t1) -- (t1H) -- (t2H) -- (t2);
\filldraw [very thick] (t1) circle (2pt);
\filldraw [very thick] (t2) circle (2pt);
\end{tikzpicture}
}

\newcommand{\DiagramGTwo}{
\begin{tikzpicture}[baseline={([yshift=-.5ex]current bounding box.center)}]
\coordinate (t) at (0,0);
\coordinate (t1) at ($1*(lengthInnerBlock)$);
\coordinate (t0) at ($1*(lengthInnerBlock)+(lengthOuterLeg)$);
\coordinate (tH) at ($(t)+(height)$);
\coordinate (t1H) at ($(t1)+(height)$);
\draw[thick] (t) -- (t0);
\draw[thick] (t) -- (tH) -- (t1H) -- (t1);
\filldraw [very thick] (t) circle (2pt);
\filldraw [very thick] (t1) circle (2pt);
\end{tikzpicture}
}

\newcommand{\DiagramGFourOne}{
\begin{tikzpicture}[baseline={([yshift=-.5ex]current bounding box.center)}]
\coordinate (t) at (0,0);
\coordinate (t1) at ($1*(lengthInnerBlock)$);
\coordinate (t2) at ($2*(lengthInnerBlock)$);
\coordinate (t3) at ($3*(lengthInnerBlock)$);
\coordinate (t0) at ($3*(lengthInnerBlock)+(lengthOuterLeg)$);
\coordinate (tH) at ($(t)+(height)$);
\coordinate (t1H) at ($(t1)+(smallHeight)$);
\coordinate (t2H) at ($(t2)+(height)$);
\coordinate (t3H) at ($(t3)+(smallHeight)$);
\draw[thick] (t) -- (t0);
\draw[thick] (t) -- (tH) -- (t2H) -- (t2);
\draw[thick] (t1) -- (t1H) -- (t3H) -- (t3);
\filldraw [very thick] (t) circle (2pt);
\filldraw [very thick] (t1) circle (2pt);
\filldraw [very thick] (t2) circle (2pt);
\filldraw [very thick] (t3) circle (2pt);
\end{tikzpicture}
}

\newcommand{\DiagramGFourTwo}{
\begin{tikzpicture}[baseline={([yshift=-.5ex]current bounding box.center)}]
\coordinate (t) at (0,0);
\coordinate (t1) at ($1*(lengthInnerBlock)$);
\coordinate (t2) at ($2*(lengthInnerBlock)$);
\coordinate (t3) at ($3*(lengthInnerBlock)$);
\coordinate (t0) at ($3*(lengthInnerBlock)+(lengthOuterLeg)$);
\coordinate (tH) at ($(t)+(height)$);
\coordinate (t1H) at ($(t1)+(smallHeight)$);
\coordinate (t2H) at ($(t2)+(smallHeight)$);
\coordinate (t3H) at ($(t3)+(height)$);
\draw[thick] (t) -- (t0);
\draw[thick] (t) -- (tH) -- (t3H) -- (t3);
\draw[thick] (t1) -- (t1H) -- (t2H) -- (t2);
\filldraw [very thick] (t) circle (2pt);
\filldraw [very thick] (t1) circle (2pt);
\filldraw [very thick] (t2) circle (2pt);
\filldraw [very thick] (t3) circle (2pt);
\end{tikzpicture}
}

\newcommand{\DiagramGFourThree}{
\begin{tikzpicture}[baseline={([yshift=-.5ex]current bounding box.center)}]
\coordinate (t) at (0,0);
\coordinate (t1) at ($1*(lengthInnerBlock)$);
\coordinate (t2) at ($2*(lengthInnerBlock)$);
\coordinate (t3) at ($3*(lengthInnerBlock)$);
\coordinate (t0) at ($3*(lengthInnerBlock)+(lengthOuterLeg)$);
\coordinate (tH) at ($(t)+(height)$);
\coordinate (t1H) at ($(t1)+(height)$);
\coordinate (t2H) at ($(t2)+(height)$);
\coordinate (t3H) at ($(t3)+(height)$);
\draw[thick] (t) -- (t0);
\draw[thick] (t) -- (tH) -- (t1H) -- (t1);
\draw[thick] (t2) -- (t2H) -- (t3H) -- (t3);
\filldraw [very thick] (t) circle (2pt);
\filldraw [very thick] (t1) circle (2pt);
\filldraw [very thick] (t2) circle (2pt);
\filldraw [very thick] (t3) circle (2pt);
\end{tikzpicture}
} \title{
	The connection between time-local and time-nonlocal perturbation expansions
}
\author{K. Nestmann$^{(1,2)}$}
\author{M. R. Wegewijs$^{(1,2,3)}$}
\affiliation{
  (1) Institute for Theory of Statistical Physics,
      RWTH Aachen, 52056 Aachen,  Germany
  \\
  (2) JARA-FIT, 52056 Aachen, Germany
  \\
  (3) Peter Gr{\"u}nberg Institut,
  Forschungszentrum J{\"u}lich, 52425 J{\"u}lich,  Germany
}
\pacs{
}
 
\begin{abstract}

There exist two canonical approaches
to describe open quantum systems
by a time-evolution equation:
the Nakajima-Zwanzig quantum master equation, featuring a time-nonlocal memory kernel $\K$, and the time-convolutionless equation with a time-local generator $\G$.
These key quantities have recently been shown to be connected by an exact \emph{fixed-point} relation
[Phys. Rev. X 11, 021041 (2021)].
Here we show that this implies a \emph{recursive} relation between their perturbative expansions, allowing a series for the kernel $\K$ to be translated directly into a corresponding series for the more complicated generator $\G$.
This leads to an elegant way of computing the generator using well-developed, \emph{standard} memory-kernel techniques for strongly interacting open systems. Moreover,
it	
allows for an unbiased comparison of time-local and time-nonlocal approaches independent of the particular technique chosen to calculate expansions of $\K$ and $\G$ (Nakajima-Zwanzig projections, real-time diagrams, etc.). We illustrate this for leading and next-to-leading order calculations of $\K$ and $\G$ for the single impurity Anderson model using both the bare expansion in the system-environment coupling  and a more advanced \emph{renormalized} series.
We compare the different expansions obtained, quantify the legitimacy of the generated dynamics (complete positivity) and benchmark with the exact result in the non-interacting limit.
 \end{abstract}

\maketitle
\section{Introduction\label{sec:intro}}

In contrast to closed quantum systems, which are always described by the Schr\"{o}dinger equation, a peculiar feature of
the description of open quantum systems is that one has to \emph{choose} between
two equivalent exact quantum master equations.
On the one hand, there is the \emph{time-nonlocal} quantum master equation (Nakajima-Zwanzig), producing the evolution map~\footnote{We consider dynamics generated by a unitary evolution of a system which is not correlated initially with its environment.} $\rho(t)=\Pi(t,t_0)\rho(t_0)$ by convolution with a memory kernel~$\K$:
\begin{align}
\tfrac{d}{dt}\Pi(t,t_0) = -i \int_{t_0}^t ds \K(t,s) \Pi(s,t_0).
\label{eq:qme-nonlocal}
\end{align}
The \emph{time-local} quantum master equation, on the other hand, features a time-dependent generator $\G$
\begin{align}
\tfrac{d}{dt}\Pi(t,t_0) = -i \G(t,t_0) \Pi(t,t_0).
\label{eq:qme-local}
\end{align}
The time-nonlocal memory kernel $\K$ is closely connected to the microscopic point of view and perturbation expansions, which are at the focus of this paper. It captures the retarded effect of the environment on the system (memory). Several methods have been developed to compute $\K$, even for complicated models featuring strong interaction and memory effects \cite{Koenig96a,Koenig96b,Pedersen07, Schoeller09, Koller10, Kern13, Schoeller18,Cohen11, Pletyukhov12a, Lindner18, Lindner19}. The opposite point of view predominant in quantum information and dynamics is of independent interest, where one starts from given general properties of the dynamical map $\Pi$, which are more easily related to the generator $\G$.
However, the computation of the time-local generator is much more challenging. Nevertheless this has been approached from various angles~\cite{Breuer01, Timm11, Karlewski14, Ferguson21, Mazza21}
motivated, for example, by the fact that $\G$ is the quantity of choice for understanding the \enquote{(non-)Markovianity}~\cite{Hall14,Breuer16rev,Benatti17,Chruscinski18} of the dynamics by its connection to both P-~\cite{Wissmann15,Bae2016} and CP-divisibility~\cite{Rivas10,Chruscinski12a,Rivas14}. This seems practically impossible to achieve with a description of the dynamics based on $\K$~\cite{Chruscinski16,Chruscinski17b,Filippov18}.
Also, the widely applied stochastic simulation methods~\cite{Wiseman_book} are also invariably based on $\G$~\cite{Piilo08,Caiaffa17,Smirne20}.
Finally, it is only using $\G$ that one can directly study geometric and possible topological phases in open quantum systems~\cite{Sarandy05,Sarandy06,Li14, Krimer19,Riwar19}, which emerge, for example, in the study of pumping \cite{Splettstoesser06,Sinitsyn09,Pluecker17}.

In \Ref{Nestmann21a} we recently found that $\G$ is a \emph{fixed point} of a Laplace-like transformation functional induced by $\K$, in short $\G=\hat\K[\G]$, and related work was reported in \Refs{Smirne10,Vacchini10,Megier20}. Besides revealing surprising exact relations between spectral properties of $\G$, $\K$ and $\Pi$ this result makes it possible to compute $\G$ from $\K$ without first constructing the full propagator $\Pi$.
In the present paper we apply this relation, reviewed in \Sec{sec:functional},
to solve another open problem in the field of open system dynamics:
We show it can be used to translate a given perturbation series for $\K$ into a corresponding series for the more complicated $\G$.
Importantly, approximations to $\K$ and $\G$ computed to the \emph{same} order using the \emph{same} perturbative scheme give \emph{different} approximate evolutions $\Pi$ due to the difference in time (non-)locality of Eqs.~\eq{eq:qme-nonlocal}--\eq{eq:qme-local}.
It is an intriguing and delicate question which of the two expansions does \enquote{better}.	
This issue calls for a general way to meaningfully compare such expansions.
In \Sec{sec:anderson} we illustrate how this can be done both analytically and numerically using the interacting Anderson quantum dot
as an example by calculating both $\K$ and $\G$
using a particular diagrammatic technique~\cite{Schoeller09, Saptsov14}.
We stress that this technique is not at all essential to our central result. The same result is obtained,
for instance, using the Nakajima-Zwanzig projection technique instead.
Moreover, we are also flexible in the choice of the expansion reference,
allowing us to explore another, \emph{renormalized} perturbation theory,
to extend the potential usefulness of our results in regimes of strong coupling.
This more powerful expansion was initially developed for $\K$ as first stage of a continuous RG-flow method for open quantum systems~\cite{Schoeller09,Andergassen11a, Kashuba13} to deal with strong dissipative coupling and non-equilibrium.
It was later studied on its own merits~\cite{Saptsov12,Saptsov14,Schulenborg16,Reimer19b} and revealed powerful exact relations~\cite{Schulenborg16,Bruch21a}. However, applications to the \emph{transient} time-evolution of interacting systems analyzed in the present work were not yet explored.

 \section{Fixed-point equation connecting generator and memory-kernel\label{sec:functional}}

Given a memory kernel $\K$ (or an approximation to it), we ask the question how to construct a corresponding (approximation to the) generator $\G$. One way to achieve this is to first solve \Eq{eq:qme-nonlocal} for the evolution $\Pi$, from which one can then reverse engineer the generator
\begin{align}
-i \G(t,t_0) =  \big[ \tfrac{d}{dt}\Pi(t,t_0) \big] \Pi(t,t_0)^{-1}
\label{eq:G_from_Pi}
.
\end{align}
Interestingly,
it was shown in \Ref{Nestmann21a} that it is also possible to compute $\G$ from $\K$ \emph{without} solving for the complete evolution first. To do so,
for the given kernel $\K$ one considers	
the functional $\hat\K[X]$, which maps a time-dependent superoperator $X(t,t_0)$ to another such object,
\begin{align}
\hat{\K}[X](t,t_0) \coloneqq \int_{t_0}^t ds
\,  \K(t,s) 
\T_{\rightarrow} e^{ i\int_s^t d\tau X(\tau,t_0)}
\label{eq:functional_def}
,
\end{align}
where $\T_{\rightarrow}$ denotes anti-time-ordering. It then follows that $\G$ is a fixed point of this functional~\cite{Nestmann21a},
\begin{align}
\G(t,t_0) = \hat{\K}[\G](t,t_0)
\label{eq:fixed_point}
.
\end{align}
One possible way to use this for computing $\G$ is to first find a reasonable initial guess for the generator denoted $\G_{0}$, and then to iterate the functional $i=1,2,3,\dots$ times,
\begin{align}
\G_i(t,t_0) \coloneqq \hat{\K}\left[\G_{i-1}\right](t,t_0)
\label{eq:fixed_point_iteration}
.
\end{align}
It was shown for specific models that this iteration converges with $i \rightarrow \infty$ to the exact generator, and that this is even possible if $\G$ has physical singularities at isolated times~\cite{Nestmann21a}. A mathematical analysis of the success of this strategy  is certainly interesting and should address questions of uniqueness and stability of the fixed point raised in \Ref{Nestmann21a}. In this paper we instead focus on useful
formal implications of the fixed-point equation \eq{eq:fixed_point}, in particular, how it leads to a natural reorganization of perturbation expansions \emph{when the time-(non)locality} of the quantum master equation is \emph{altered}.

Thus, the goal is to find a perturbative expansion for $\G$ based on a corresponding expansion for $\K$.
By \enquote{corresponding} we mean that both series count powers of the same formal expansion parameter.
We start by decomposing $\K$ as
\begin{align}
\K(t,s) = \K_L \bar{\delta}(t-s) + \K_N(t,s)
\label{eq:decomposition_K}
.
\end{align}
with a $\bar{\delta}$ distribution normalized as $\int_0^\infty ds f(s) \bar{\delta}(t-s) = f(t)$. Here $\K_L$ determines
a time-local
part, which at first is taken to be the  uncoupled system Liouvillian, and $\K_N$ denotes the remaining environment part
due to non-zero coupling,	
which contains time-nonlocal contributions.
The kernel $\K^{(0)}(t,s)=\K_L\bar{\delta}(t-s)$ producing semigroup dynamics $\Pi^{(0)}\coloneqq e^{ -i \K_L t}$ via \Eq{eq:qme-nonlocal} will be the reference point of the perturbation theory.
Importantly, we will also allow for a renormalized expansion, in which a \emph{further} time-local contribution
--which was still contained in $\K_N$--
is included in $\K_L$, making the reference evolution $\Pi^{(0)}$ \emph{dissipative},
see details below [\Eq{eq:extract}].
In either case we assume for simplicity that $\K_L$ is time-independent, but this is not a limiting assumption.

Decomposing $\G(t,t_0) = \K_L + \G_N(t,t_0)$ analogously, the fixed point equation \eq{eq:fixed_point} implies that $\G_N$ obeys
\begin{align}
\G_N(t,t_0) = \int_{t_0}^t ds
\,  \K_N(t,s) 
\T_{\rightarrow} e^{ i\int_s^t d\tau \left[\K_L + \G_N(\tau,t_0)\right]}
\label{eq:fixed_point_G_N}
.
\end{align}
We correspondingly use $\G^{(0)}=\K_L$ as a reference for the expansion of $\G$.
Assuming that the nonlocal part of the memory kernel is given by a series
in some formal parameter,
$\K_N = \K^{(1)} + \K^{(2)} + \dots,$
we can derive the \emph{corresponding} series for $\G_N=\G^{(1)}+\G^{(2)}+\dots$ in the \emph{same} parameter by first expanding the anti-time-ordered exponential in \Eq{eq:fixed_point_G_N} and then matching orders. The first two terms explicitly read
\begin{align}
\G^{(1)}(t,t_0) =& \int_{t_0}^t ds \, \K^{(1)}(t,s) e^{ -i \K_L (s-t)}, \label{eq:G1_from_K} \\
\G^{(2)}(t,t_0) =& \int_{t_0}^t ds \, \K^{(2)}(t,s) e^{ -i \K_L (s-t)}  \notag \\
              &+ i \int_{t_0}^t ds \int_{s}^t d\tau \, \K^{(1)}(t,s) \notag \\
               &\quad \times e^{ -i \K_L (s-\tau)} \G^{(1)}(\tau,t_0) e^{ -i \K_L (\tau-t)}  \label{eq:G2_from_K}.
\end{align}
The general $n$-th order $\G^{(n)}$ is similarly given by
\begin{widetext}
\begin{align}
\G^{(n)}(t) = \sum_{l=0}^{n-1} i^l \sum_{\Sigma_i m_i = n} \underset{t_0 < \tau_0 < \dots < \tau_{l} < t}{\int d\tau_0 \dots d\tau_{l}} \K^{(m_0)}(t,\tau_0) e^{ -i \K_L (\tau_0-\tau_1)} \G^{(m_1)}(\tau_1) e^{ -i \K_L (\tau_1-\tau_2)} \G^{(m_2)}(\tau_2) \cdots \G^{(m_{l})}(\tau_{l}) e^{ -i \K_L (\tau_{l}-t)}
\label{eq:total_G_from_K}
,
\end{align}
\end{widetext}
where the second sum runs over $m_0,\dots,m_l>0$. We thus see that the \emph{fixed-point} equation automatically organizes the series expansion of $\G_N$ into a \emph{recursive} form. It it well known that when computing the memory kernel $\K$, for example using standard projection operator or diagrammatic techniques, one obtains only time ordered contributions (convolutions), whereas the generator $\G$ has a more complicated structure involving combinations of non-time ordered integrations. The recursive reorganization implied by the fixed-point relation completely disentangles this nontrivial structure: \Eq{eq:total_G_from_K} reveals that collecting all \emph{time-ordered} contributions one obtains precisely the various memory kernel components $\K^{(n)}$ -- obtainable by well-developed
\emph{standard}	
 techniques -- and that the remaining integrations are exclusively anti-time ordered.
In the remainder of the paper we will exploit this insight
using the diagrammatic approach, noting that one may equivalently use the projection operator technique.

So far we refrained from making use of the propagator. However, if the orders of $\Pi$ are formally known this can be useful. By similarly expanding $\Pi=\Pi^{(0)}+\Pi^{(1)}+\dots$ and inserting into $\G_N \Pi = \K_N * \Pi$
[\Eq{eq:qme-local} $= $ \Eq{eq:qme-nonlocal}],	
written as
$	\G_N = \left[ \K_N * \Pi - \G_N \left(\Pi - e^{-i \K_L t} \right) \right]  e^{i \K_L t}
$,
one obtains a useful reorganization,
expressing the $n$-th order of $\G$ in terms of its lower orders with the help of both the memory kernel $\K$ and the propagator up to order $n$,
\begin{align}
	\G^{(n)} =& \K^{(n)} * \Pi^{(0)} e^{i \K_L t} \label{eq:G_orders_pretty} \\
	         & + \sum_{j=1}^{n-1} \left[ \K^{(n-j)} * \Pi^{(j)} - \G^{(n-j)} \cdot \Pi^{(j)} \right] e^{ i \K_L t}  \notag
	.
\end{align}
The above relations are key results of this paper
and apply generally to open quantum system.

Since the fixed-point equation
is flexible and
can be exploited in various ways, it is important to keep the following in mind.
We are interested
here	
in comparing different solutions generated by corresponding perturbative expansions, the difference arising from their time-(non)locality.
We want to explore whether summing up partial contributions in a time-local framework leads	to better results in some sense than when doing the corresponding sum in the time-nonlocal framework.
Given, for example, a second order approximation to the kernel, $\K_\text{pert} = \K_L \bar{\delta} + \K^{(1)}+ \K^{(2)}$ ,
it is only meaningful to compare the evolution it produces via \Eq{eq:qme-nonlocal} with the evolution produced via \Eq{eq:qme-local} by the \emph{perturbative} $\G_\text{pert} \approx \K_L + \G^{(1)}+ \G^{(2)}$, where we in both cases expand in the same parameter.
When taking $\K_L=L$, this corresponds to contrasting the well-established bare perturbation expansions of $\K$~\cite{Feynman63,Koenig96a,Koenig96b} and $\G$~\cite{Tokuyama75,Tokuyama76,Shibata77,Shibata80,Chaturvedi79,Breuer01,BreuerPetruccione} whose traditional derivations are very difficult to compare.
Since we are able to treat both expansions in the same, standard way
a comparison becomes possible.

As a remark, we point out another way of exploiting the fixed-point equation.
It namely defines an approximate \emph{self-consistent} generator $\G_\text{sc}$.
which produces exactly the \emph{same} evolution
via \Eq{eq:qme-local} as the perturbative $\K_\text{pert}$ does via \Eq{eq:qme-nonlocal}.
In other words~\footnote{
	As pointed out already in \Ref{Nestmann21a},
	the fixed point equation merely expresses the relation between a kernel and a generator of the same evolution, even when that evolution is an approximation.
},
it satisfies the fixed-point equation
$\G_\text{sc}= \hat{\K}_\text{pert}[ \G_\text{sc}]$
being self-consistent relative to the kernel approximation.	
This approximate but self-consistent generator \emph{differs} from the one considered in this paper,
$\G_\text{sc} \neq \G_\text{pert}$.
One should realize that when using $\G_\text{sc}$ one essentially gives up calculating the generator directly
but formulates all approximations using $\K$ and afterwards produces the \emph{equivalent} generator
(as opposed to \emph{corresponding}).
The advantage of the fixed-point equation is that one achieves this without first constructing the solution $\Pi_\text{pert}$ from $\K_\text{pert}$,
by a self-consistent iterative computation which was numerically explored in \Ref{Nestmann21a}
for an exactly known $\K$.
The self-consistent generator $\G_\text{sc}$ is of independent interest, but beyond the present scope
and will not be considered here.
We merely note that this is useful,
since it allows, for example, to determine whether the evolution generated by an approximate kernel $\K_\text{pert}$ is Markovian in either the P- or CP-divisible sense, since this can only be decided by inspection of the \emph{equivalent} time-local generator $\G_\text{sc}$ written in the Gorini–Kossakowski–Sudarshan–Lindblad (GKSL) form~\cite{Rivas10,Rivas14,Wissmann15}.
As another example, with $\G_\text{sc}$ in hand one can investigate individual quantum trajectories of \emph{approximations} to non-Markovian dynamics using well-developed stochastic simulation techniques~\cite{Piilo08,Caiaffa17,Smirne20} based on the \emph{time-local} QME,
avoiding the memory-integrals required by time-nonlocal stochastic simulation methods~\cite{Diosi98,Strunz99}.

 \section{Transient behavior of the single impurity Anderson model\label{sec:anderson}}

\subsection{Anderson model in Liouville space}

As an example, we consider a single orbital quantum dot with spin described by
\begin{align}
H=\epsilon (n_\uparrow + n_\downarrow) + U n_\uparrow n_\downarrow.
\end{align}
Here $\epsilon$ is the energy of the orbital, $n_\sigma = d^\dagger_\sigma d_\sigma$ the number operator for spin $\sigma$ and $U$ is the Coulomb interaction. This quantum dot is connected to several free electron reservoirs
\begin{align}
H_R= \sum_{r \sigma} \int d \omega (\omega + \mu_r) a^\dagger_{r \sigma}(\omega) a_{r \sigma}(\omega).
\end{align}
We allow that
the reservoirs labeled by $r$ are initially in thermal equilibrium at different temperatures $T_r$ and chemical potentials $\mu_r$
but in illustrations we focus on $T_r=T$.
The tunnel junctions are modeled with the Hamiltonian
\begin{align}
H_T= \sum_{r \sigma} \int d\omega \sqrt{\frac{\Gamma_{r \sigma}}{2\pi}} \left( d_\sigma a^\dagger_{r \sigma}(\omega) + a_{r \sigma}(\omega) d_\sigma^\dagger \right),
\label{eq:H_tunnel}
\end{align}
where $\Gamma_{r \sigma}$ is the real-valued, spin-dependent spectral density of reservoir $r$
assumed to be energy independent (wideband limit). Thus the total Hamiltonian is
\begin{align}
H_{\text{tot}} = H + H_R + H_T.
\end{align}

For the calculation of the open system dynamics a formalism based on superoperators is convenient.
Here we use the superfermion approach to Liouville space introduced in \Ref{Schoeller09} using the later formulation of \Ref{Saptsov14}, where details and comparison with other constructions can be found. Defining first the shorthand notation
\begin{align}
d_{\eta \sigma} \coloneqq \left\{\begin{array}{lrl}
d^\dagger_{\sigma} & \text{for} & \eta = +\\
d_{\sigma} & \text{for} & \eta = -
\end{array}\right.
,
\end{align}
the superfermions are superoperators defined as
\begin{align}
G^p_{\eta \sigma} \bullet \coloneqq \frac{1}{2} \left( d_{\eta \sigma} \bullet + p (-\one)^n \bullet (-\one)^n d_{\eta \sigma} \right)
\label{eq:superfermion_def}
,
\end{align}
where $p=+$ gives a creation and $p=-$ an annihilation superoperator and $(-\one)^n \coloneqq (\one - 2 n_\uparrow)(\one - 2 n_\downarrow)$ denotes the fermion parity operator. The superfermions act in the Liouville-Fock space analogously to the way that ordinary creation/annihilation operators act in the Hilbert-Fock space. The \emph{supervacuum} state $\Ket{0}$ corresponding to \Eq{eq:superfermion_def} is given by the infinite-temperature stationary state $\Ket{0} \coloneqq \tfrac{1}{2} \one$
considered as a supervector.	
From this a complete basis for the Liouville-Fock space is generated by the superfermionic creation operators ($G^+_{\eta \sigma}$) in the usual way~\cite{Saptsov12}. This choice of fields and vacuum is particularly well-adapted to the perturbation expansion, as we will see [\Eq{eq:infinite_T_Pi}]. Furthermore, the superfermions anticommute,
\begin{align}
\left\{ G_{\eta_1 \sigma_1}^{p_1}, G_{\eta_2 \sigma_2}^{p_2} \right\} = \delta_{p_1 \bar{p}_2} \delta_{\eta_1 \bar{\eta}_2} \delta_{\sigma_1 \sigma_2}
\label{eq:superfermions_anticommute}
,
\end{align}
where $\bar x \coloneqq -x$, and they satisfy a super-Pauli principle, which states that it is formally impossible to create or destroy two identical superfermions
\begin{align}
 \left( G_{\eta \sigma}^{p} \right)^2 = 0.
\end{align}
In the same fashion as ordinary operators are written down as strings of creation/annihilation field operators this can also be done for superoperators. For example, the local Liouvillian, $L\bullet \coloneqq [H,\bullet]$, is given by~\cite{Saptsov14}
\begin{align}
L =& \sum_{\eta \sigma} \Big[ \bar\eta \left( \epsilon + \tfrac{1}{2} U \right) G^{+}_{\bar\eta \sigma} G^{-}_{\eta \sigma} \Big. \\
& + \Big. \tfrac{1}{2} U \left( G^{+}_{\bar\eta \sigma} G^{-}_{\eta \sigma} G^{-}_{\bar\eta \bar\sigma} G^{-}_{\eta \bar\sigma} + G^{+}_{\bar\eta \bar\sigma} G^{+}_{\eta \bar\sigma} G^{+}_{\bar\eta \sigma} G^{-}_{\eta \sigma} \right) \Big] \notag
.
\end{align}

\subsection{Comparing approximations}

Before we start investigating different approximations, we note that one obvious 
comparison to consider is the difference of the approximate state evolution with the exact one. In the following we will  quantify the difference between two density operators $\rho$ and $\sigma$ using the trace distance
\begin{align}
	D\big( \rho, \sigma \big) \coloneqq \frac{1}{2} \Tr \sqrt{(\rho-\sigma)^\dagger (\rho-\sigma)}
	.
\end{align}
This is a metric on the space of density operators with a physical meaning: it determines the optimal probability~\cite{Helstrom69} of distinguishing $\rho$ and $\sigma$ drawn from a unbiased ensemble and also plays a central role in the study of non-Markovianity~\cite{Breuer09a,Chruscinski11,Chruscinski18}. Also, given any observable $A$ it can be shown that the trace distance bounds the difference of its expectation value using either $\rho$ or $\sigma$ relative to the largest singular value $\left\lVert A \right\rVert_\infty$ (Schatten $\infty$-norm~\cite{Wolf12}):
\begin{align}
	\big| \langle A \rangle_\rho - \langle A \rangle_\sigma \big| / \left\lVert A \right\rVert_\infty
	\leq 2  D\big( \rho, \sigma \big) 
	.
\end{align}

Of course, for problems of actual interest
the exact solution needed for this comparison is not available. 	
Then the most basic thing to check is whether the approximation stays \emph{physical}, for which two criteria need to be fulfilled. First, the trace  needs to be preserved. This is automatically guaranteed in \Eq{eq:all_K2_diagrams} and \Eq{eq:all_K4_diagrams} term by term, because on the left there is always a creation superfermion $G^+_{\eta \sigma}$, which has the trace functional as a left zero eigenvector~\cite{Saptsov14}. Second, an evolution needs to be \emph{completely positive} (CP). This means that when the evolution is applied to the system entangled with any auxilliary system, the composite output state is still valid. It is well known how to check this based on the \emph{solution} for the \emph{propagator} $\Pi(t)$ by checking the positivity of the so-called Choi operator~\cite{Wolf12},
but frequently this	is
not discussed in studies of advanced approximation strategies which go beyond the applicability of the GKSL theorem~\cite{GKS76,Lindblad76}.
One should note that CP can \emph{not} be determined by looking at a trajectory $\rho(t)=\Pi(t)\rho_0$ starting from some specific initial state $\rho_0$, see \Ref{Reimer19b} for details and examples. Moreover, an evolution $\Pi(t)$ may even produce valid quantum states $\rho(t)$ for any valid input state $\rho_0$,
i.e., be positivity preserving,
but \emph{still fail} to be completely positive. This is not a rare situation and such unphysical maps are well known from their mathematical application to the detection of entanglement~\cite{Horodecki09}.

\subsection{Bare perturbation theory \label{sec:bare_perturbation}}

\subsubsection{Bare perturbation theory for kernel $\K$}

One systematic way of computing $\K$ is by a bare perturbation expansion in the coupling to the environment. This is substantially simplified~\cite{Leijnse08,Emary10,Emary11,Saptsov12} by combining diagrammatic~\cite{Koenig96b,Koller10} and Liouville-Fock space techniques~\cite{Schoeller09,Schulenborg16,Schoeller18} and exploiting the wideband limit from the very beginning.
We refer the reader to \Ref{Saptsov14} for further details.

Because the total Hamiltonian $H_{\text{tot}}$ is time-independent, the kernel only depends on the difference of its time arguments, $\K(t,s)=\K(t-s)$. Using the decomposition \eq{eq:decomposition_K} we have $\K_L=L$ and the goal is to compute the nonlocal part $\K_N=\K^{(1)} + \K^{(2)} + \dots$, where each term $\K^{(n)}$ contains $n$ tunneling contributions [\Eq{eq:H_tunnel}]. Because of the bilinear structure of $H_T$ it follows that all odd orders vanish. The first two nonvanishing orders are then diagrammatically represented by
\begin{align}
-i\K^{(2)}(t) =& \DiagramKTwo, \label{eq:all_K2_diagrams} \\
-i\K^{(4)}(t) =& \DiagramKFourTwo+\DiagramKFourOne. \label{eq:all_K4_diagrams}
\end{align}
The diagrams are specifically given by
\begin{align}
\DiagramKTwo =& -\sum_{p \eta \sigma}\gamma^p_{\eta \sigma}(t) G^+_{\eta \sigma} e^{-i L t} G^{\bar p}_{\bar\eta \sigma}, \label{eq:diagram_K_2} \\
\DiagramKFourTwo =& \sum_{p_1 \eta_1 \sigma_1} \sum_{p_2 \eta_2 \sigma_2} \int_0^t dt_1 \int_0^{t_1} dt_2 \notag \\
& \mkern-125mu \gamma^{p_1}_{\eta_1 \sigma_1}(t)  \gamma^{p_2}_{\eta_2 \sigma_2}(t_1 - t_2) \label{eq:diagram_K_4_2} \\
& \mkern-125mu \times  G^+_{\eta_1 \sigma_1} e^{-i L (t-t_1)} G^+_{\eta_2 \sigma_2} e^{-i L (t_1-t_2)} G^{\bar{p}_2}_{\bar\eta_2 \sigma_2} e^{-i L t_2} G^{\bar{p}_1}_{\bar\eta_1 \sigma_1}, \notag \\
\DiagramKFourOne =& - \sum_{p_1 \eta_1 \sigma_1} \sum_{p_2 \eta_2 \sigma_2} \int_0^t dt_1 \int_0^{t_1} dt_2 \notag \\
& \mkern-125mu \gamma^{p_1}_{\eta_1 \sigma_1}(t-t_2)  \gamma^{p_2}_{\eta_2 \sigma_2}(t_1) \label{eq:diagram_K_4_1} \\
& \mkern-125mu \times G^+_{\eta_1 \sigma_1} e^{-i L (t-t_1)} G^+_{\eta_2 \sigma_2} e^{-i L (t_1-t_2)} G^{\bar{p}_1}_{\bar\eta_1 \sigma_1} e^{-i L t_2} G^{\bar{p}_2}_{\bar\eta_2 \sigma_2} \notag
.
\end{align}
Here the contraction functions read
\begin{align}
\gamma^{p}_{\eta \sigma}(t) = \left\{\begin{array}{lrl}
\tfrac{1}{2}\sum_r \Gamma_{r \sigma} \bar\delta(t) & \text{for} & p = +\\
-i \sum_r \dfrac{\Gamma_{r \sigma} T_r}{\sinh(t T_r \pi)} e^{i \bar\eta \mu_r t} & \text{for} & p = -
\end{array}\right.
\label{eq:contractions}
.
\end{align}
The $\bar\delta$ distribution in the time-local $\gamma^{+}_{\eta \sigma}$ contraction arises due to the wideband limit, which was already incorporated into the definition of the Hamiltonians. As another consequence of this, the time-nonlocal $\gamma^{-}_{\eta \sigma}$ contractions contains a singularity at $t=0$.
Importantly, in \App{app:everything_finite} we show that the special algebra of the superfermions elegantly ensures that all shown diagrams stay finite nevertheless. There it is explained how the contributions combine to yield convergent time integrals, which can be straightforwardly implemented.

\begin{figure*}[t]
	\subfloat{
		\centering
		\includegraphics[width=0.45\linewidth]{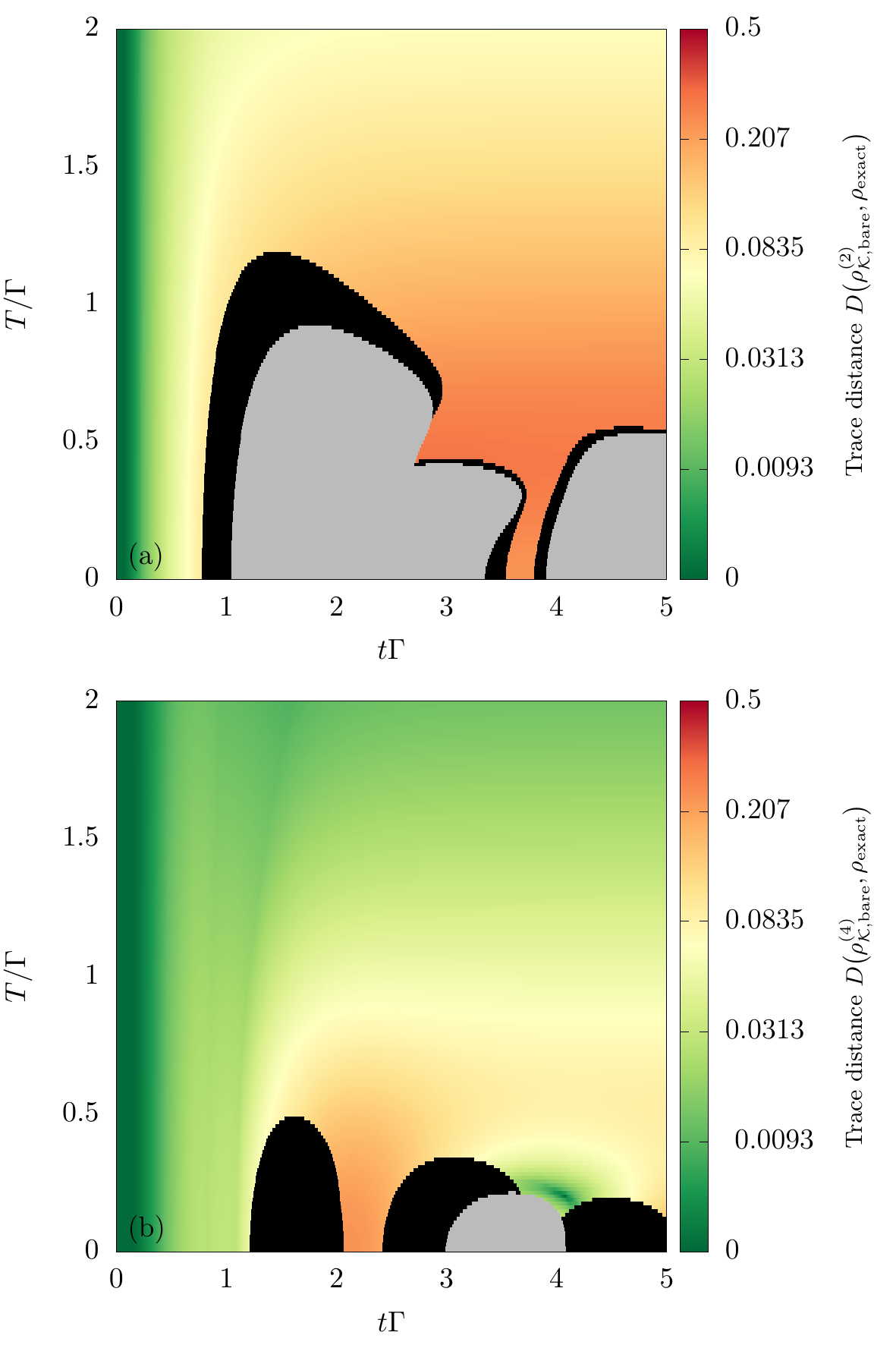}}
	\subfloat{
		\centering
		\includegraphics[width=0.45\linewidth]{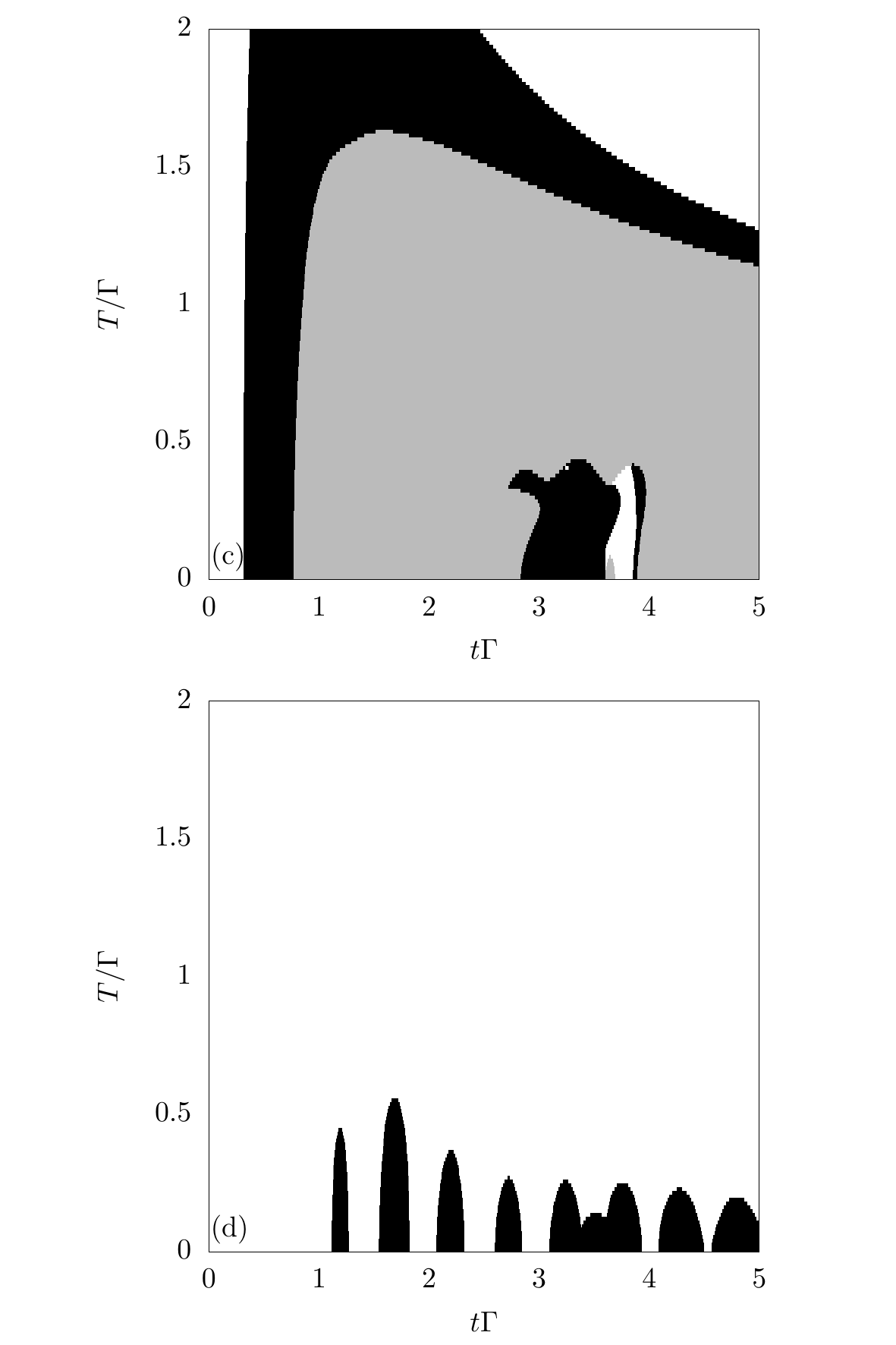}
	}
	\caption{
		Transient evolution of an initially unoccupied Anderson dot detuned by $\epsilon=2\Gamma$ connected to a left and right reservoir at the same temperatures $T_L=T_R=T$ and biased chemical potentials $\mu_L = 0$, $\mu_R=-0.2 \Gamma$. Regions in which the approximated state is not positive are shown in gray. Regions in which the approximated state is positive, but the propagator is not completely positive are shown in black.
		(a) Non-interacting case: Trace distance $D( \rho^{(2)}_{\K,\text{bare}}(t), \rho_{\text{exact}}(t) )$ between the second-order bare time-nonlocal solution $\rho^{(2)}_{\K,\text{bare}}(t)$ and the exact solution.
		(b) Non-interacting case: Trace distance $D( \rho^{(4)}_{\K,\text{bare}}(t), \rho_{\text{exact}}(t) )$.
		(c)-(d) Interacting case $U=10\Gamma$: regions of (complete) positivity for (c) $\rho^{(2)}_{\K,\text{bare}}$ and (d) for $\rho^{(4)}_{\K,\text{bare}}$.
	}\label{fig:bare_nz_vs_exact}
\end{figure*}

We illustrate the above for a generic set of parameters in \Fig{fig:bare_nz_vs_exact}(a--b). There we solve the time-nonlocal equation~\eq{eq:qme-nonlocal} using the numerically computed second and fourth order kernels. 
Referring to the solutions as $\rho^{(2)}_{\K,\text{bare}}(t)$ and $\rho^{(4)}_{\K,\text{bare}}(t)$ respectively, we plot their trace distance
to the exact solution $\rho_{\text{exact}}(t)$
for the noninteracting case $U=0$ as function of time $t$ and temperature $T$. As expected, the quality of each approximation is improved with higher temperature and the fourth order solution $\rho^{(4)}_{\K,\text{bare}}(t)$ has a larger range of validity than $\rho^{(2)}_{\K,\text{bare}}(t)$. At small temperatures the approximations work well only for short times $t \Gamma \lesssim 1$ where the infinite-temperature contributions dominate the dynamics..
This is also reflected by the more basic check of the complete positivity (CP) of the \emph{propagator} $\Pi(t)$, which is violated in the black and gray areas. It can be seen that $\rho^{(2)}_{\K,\text{bare}}(t)$ suffers from unphysical regimes, which become smaller when going to the next order $\rho^{(4)}_{\K,\text{bare}}(t)$. Importantly, when only checking whether the specific output \emph{state} $\rho(t)=\Pi(t) \rho_0$ is unphysical (non-positive), which is the case in the gray areas, one misses that in the black areas the approximation has \emph{already} failed, because the \emph{propagator} does not handle entanglement correctly (non-CP). 
These latter regimes are thus especially dangerous in practice.

Finally, in \Fig{fig:bare_nz_vs_exact}(c--d) we show that when turning on the interaction $U$ the unphysical area of the second order solution $\rho^{(2)}_{\K,\text{bare}}(t)$ increases. This is different for $\rho^{(4)}_{\K,\text{bare}}(t)$ for the chosen parameters: whereas the detailed shape of the unphysical areas do change with interaction, the overall size does not significantly increase.

\subsubsection{Bare perturbation theory for generator $\G$ \label{sec:bare_G}}

With the orders of the kernel $\K$ in hand and decomposing $\G=\K_L+\G^{(1)}+\G^{(2)}+\dots$ as before
our key results \eq{eq:G1_from_K}--\eq{eq:G2_from_K} make it
straightforward to compute the orders of $\G$ taking $\K_L=L$. Using our recursive relation \eq{eq:G_orders_pretty} it is furthermore straightforward to infer a diagrammatic representation for $\G$ using only
standard diagrams of $\K$ and $\Pi$ and the shorthand $\Pi_0=e^{-i L t}$:
\begin{align}
-i\G^{(2)}(t) =&
\DiagramGTwo \cdot \Pi_0^{\dagger}, \label{eq:bare_G_2} \\
-i\G^{(4)}(t) =&
\DiagramGFourOne \cdot \Pi_0^{\dagger} \notag \\
& + \DiagramGFourTwo \cdot \Pi_0^{\dagger} \notag \\
& + \DiagramGFourThree \cdot \Pi_0^{\dagger} \notag \\
& - \DiagramGTwo \cdot \Pi_0^{\dagger} \cdot \DiagramPiTwo \cdot \Pi_0^{\dagger}
\label{eq:bare_G_4}
,
\end{align}
Thus, no new technique and no new diagrammatic representation are required.
We see that the general structure consists of \emph{backward} bare propagations $\Pi_0^{\dagger}$, followed by blocks of $\K$ and $\Pi$, which only propagate forward. This is not unexpected:
by the definition of the generator, $\G=i \dot\Pi\, \Pi^{-1}$, \emph{any} expansion for $\G$ 
will contain both propagations forward (from $\dot\Pi$) and backward (from $\Pi^{-1}$). This is precisely what makes the expansion of $\G$ more complicated than that of $\K$~[\Eqs{eq:all_K2_diagrams}--\eq{eq:all_K4_diagrams}]. 

\begin{figure}[t]
	\centering
	\includegraphics[width=0.9\linewidth]{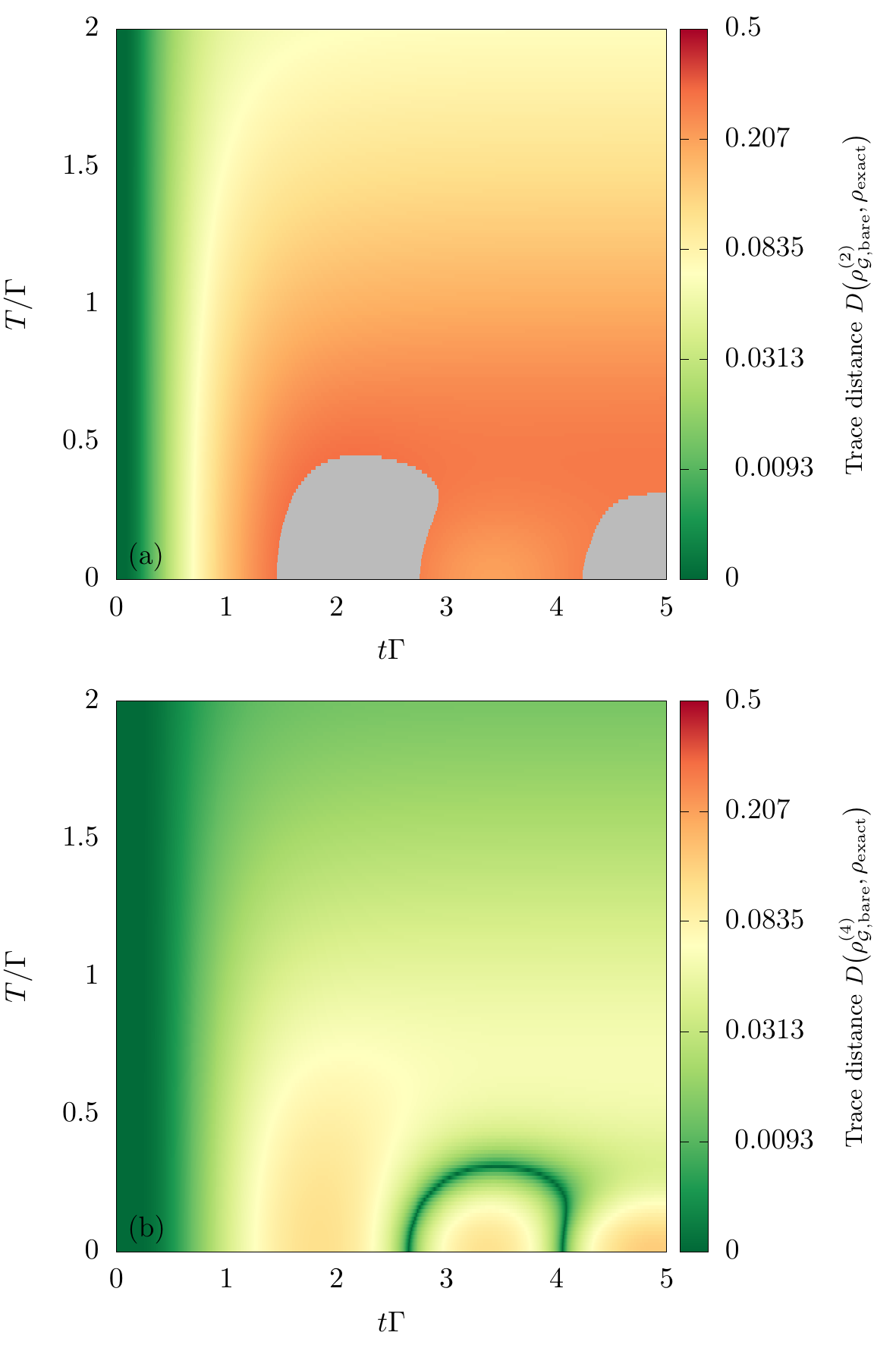}
	\caption{
		Transient evolution of an Anderson dot with the same parameters as in \Fig{fig:bare_nz_vs_exact}.
		(a) Trace distance $D( \rho^{(2)}_{\G,\text{bare}}(t), \rho_{\text{exact}}(t) )$ between the second order bare time-local solution and the exact solution.
		(b) Trace distance $D( \rho^{(4)}_{\G,\text{bare}}(t), \rho_{\text{exact}}(t) )$ between the fourth order bare time-local solution and the exact solution.
Areas in which the approximate state is not positive are shown in gray. In this case there are no areas where the approximated state is positive, but the propagator is not completely positive.
		\label{fig:bare_tcl_vs_exact}
	}
\end{figure}

In \Fig{fig:bare_tcl_vs_exact} we analyze the trace distance
of the time-local approximations	
to the exact solution in the noninteracting case $U=0$. We emphasize again that the second and fourth order time-local solutions, $\rho^{(2)}_{\G,\text{bare}}(t)$ and $\rho^{(4)}_{\G,\text{bare}}(t)$ respectively, will be different from the time-nonlocal solutions of the same order: $\rho^{(j)}_{\G,\text{bare}}(t) \neq \rho^{(j)}_{\K,\text{bare}}(t)$.

The general characteristics however stay the same: higher temperature improves the quality of the approximations and $\rho^{(4)}_{\G,\text{bare}}(t)$ outperforms $\rho^{(2)}_{\G,\text{bare}}(t)$. Compared to the time-nonlocal perturbation theory there are however considerable differences: The area where $\rho^{(2)}_{\G,\text{bare}}(t)$ is unphysical is noticeably smaller
\new{than for	
}$\rho^{(2)}_{\K,\text{bare}}(t)$. Furthermore, $\rho^{(4)}_{\G,\text{bare}}(t)$ is even physical everywhere in the plotted parameter regime. Surprisingly this is even true for strong interactions, for example of the order $U\approx 10 \Gamma$.
Notably, neither $\rho^{(2)}_{\G,\text{bare}}(t)$ nor $\rho^{(4)}_{\G,\text{bare}}(t)$ shows deceptive regimes where the state is positive, but the propagator is nevertheless not completely positive.
These observations suggest in a very basic way
that for the Anderson model the bare time-local perturbation theory is superior to the time-nonlocal one, even for the strong interaction.

\subsection{Renormalized perturbation theory\label{sec:renormalized_perturbation}}

\subsubsection{Renormalized perturbation theory for kernel $\K$ \label{sec:K}}

In \Eq{eq:contractions} the occurrence of the time-local $\bar\delta$ function in the $\gamma^{+}_{\eta \sigma}$ contraction hints at a possible simplification. Indeed it is possible to resum all these semigroup contributions systematically as was shown in \Ref{Saptsov14}.
The starting point lies in the observation that the $\gamma^{-}_{\eta \sigma}$ contraction vanishes when taking the high-temperature limit for all reservoirs, $\lim_{T_r\rightarrow \infty} \gamma^{-}_{\eta \sigma} = 0$. Then the infinite temperature kernel is exactly given by $\lim_{T_r \rightarrow \infty} \K_N(t) = \Sigma_{\infty} \, \bar\delta(t)$ with
\begin{align}
\Sigma_{\infty} \coloneqq -\frac{i}{2} \sum_{r \eta \sigma} \Gamma_{r \sigma} G^+_{\eta \sigma} G^-_{\bar\eta \sigma}
\label{eq:sigma_infinity_def}
.
\end{align}
Thus the time-nonlocal part of the kernel becomes \emph{time-local} in this limit, from which the the exact propagator can be deduced:
\begin{align}
\Pi_\infty(t) \coloneqq \lim_{T_r \rightarrow \infty} \Pi(t) = e^{-i(L + \Sigma_{\infty}) t}.
\label{eq:infinite_T_Pi}
\end{align}

Now the main idea is to set up a perturbation theory around the infinite temperature limit. Here the choice of the supervacuum $\Ket{0}$ and the associated fields $G^p_{\eta \sigma}$ provide the key advantage as follows: One makes two simple changes to the bare perturbation theory of Eqs. \eq{eq:diagram_K_2}-\eq{eq:diagram_K_4_2}~\cite{Saptsov14}. First, all Liouvillians are replaced by
\begin{align}
L \rightarrow L_\infty \coloneqq L + \Sigma_{\infty}
.
\label{eq:extract}
\end{align}
By this step one extracts in \Eq{eq:decomposition_K} a time-local part $\Sigma_{\infty}$ from the reservoir induced dynamics $\K_N$ and incorporates this into $\K_L$, leaving their sum $\K$ unaltered.	
Second, $\gamma^{+}_{\eta \sigma}$ contractions or, equivalently, $G^-_{\eta \sigma}$ vertices are no longer allowed. Thus, in the renormalized perturbation theory we have $\K_L = L_\infty$ and
\begin{subequations}
\begin{align}
\DiagramKTwo =& -\sum_{\eta \sigma}\gamma^-_{\eta \sigma}(t) G^+_{\eta \sigma} e^{-i L_\infty t} G^{+}_{\bar\eta \sigma}, \label{eq:diagram_K_2_ren} \\
\DiagramKFourTwo =& \sum_{\eta_1 \sigma_1} \sum_{\eta_2 \sigma_2} \int_0^t dt_1 \int_0^{t_1} dt_2 \notag \\
& \mkern-125mu \gamma^{-}_{\eta_1 \sigma_1}(t)  \gamma^{-}_{\eta_2 \sigma_2}(t_1 - t_2) G^+_{\eta_1 \sigma_1} e^{-i L_\infty (t-t_1)} G^+_{\eta_2 \sigma_2} \notag \\
& \mkern-125mu \times e^{-i L_\infty (t_1-t_2)} G^{+}_{\bar\eta_2 \sigma_2} e^{-i L_\infty t_2} G^{+}_{\bar\eta_1 \sigma_1}, \label{eq:diagram_K_4_2_ren} \\
\DiagramKFourOne =& - \sum_{\eta_1 \sigma_1} \sum_{\eta_2 \sigma_2} \int_0^t dt_1 \int_0^{t_1} dt_2 \notag \\
& \mkern-125mu \gamma^{-}_{\eta_1 \sigma_1}(t-t_2)  \gamma^{-}_{\eta_2 \sigma_2}(t_1) G^+_{\eta_1 \sigma_1} e^{-i L_\infty (t-t_1)} G^+_{\eta_2 \sigma_2} \notag \\
& \mkern-125mu \times e^{-i L_\infty (t_1-t_2)} G^{+}_{\bar\eta_1 \sigma_1} e^{-i L_\infty t_2} G^{+}_{\bar\eta_2 \sigma_2} \label{eq:diagram_K_4_1_ren}
.
\end{align}
\label{eq:diagrams_K_ren}
\end{subequations}

Notably, the renormalized perturbation theory is at the same time more powerful \emph{and} simpler than the original one: since $\gamma^+$ contractions are no longer allowed, there are considerably fewer terms in the renormalized perturbation theory that need to be computed. Moreover, for
vanishing interaction $U=0$ it can be shown that the renormalized series terminates and the terms~\eq{eq:diagram_K_2_ren}--\eq{eq:diagram_K_4_1_ren} already give the \emph{exact} kernel for the Anderson dot~\cite{Saptsov14}.

Compared to the bare perturbation theory the intermedite propagations between vertices are damped on a timescale of the bare tunnel rate $\sim\Gamma^{-1}$ [\Eq{eq:sigma_infinity_def}], which leads to improved convergence in the time integrations. Since this is the largest rate of decay, the higher order corrections of the renormalized perturbation theory are needed for smaller rates, i.e., they must effectively suppress decay. One thus expects that in the lower orders of this perturbation theory the oscillations described by $L$ are damped.

For the explored parameters,	
we find that even for strong interactions the renormalized fourth order solution $\rho^{(4)}_{\K,\text{ren}}$ always stays physical (same parameters as in \Fig{fig:bare_and_ren_nz}~(d), data not shown). This is however not the case for the second order renormalized solution $\rho^{(2)}_{\K,\text{ren}}$, which 
becomes	
unphysical at low temperatures, even for $U=0$ (data not shown). Larger interaction has a negative impact on the positivity of $\rho^{(2)}_{\K,\text{ren}}$.

In \Fig{fig:bare_and_ren_nz}~(a,e,f) we see that overall
compared to the bare $\K$ perturbation theory
the renormalized version replaces oscillatory behavior occurring off resonance ($\epsilon \gtrsim \Gamma$) at low $T \lesssim \Gamma$ by a rapid decay to the stationary value. We discuss the details in the next section.

\subsubsection{Renormalized perturbation theory for generator $\G$\label{sec:ren_G}}

We now \new{set up} a corrresponding approach for the time-local generator, which to our knowledge has not been explored yet.
The renormalized expansion of $\K$ can be translated to the generator $\G$ using the same steps as for the bare $\G$ expansion [\Eqs{eq:bare_G_2}--\eq{eq:bare_G_4}]. Compared to \Eqs{eq:bare_G_2}--\eq{eq:bare_G_4} one now uses the renormalized $\K$ and $\Pi$ diagrams, where only $G^+_{\eta \sigma}$ vertices are allowed
and the renormalization $L \rightarrow L_\infty = L + \Sigma_{\infty}$ is made.
Importantly, one thus also needs to replace the \emph{backward} evolutions by
\begin{align}
\Pi_0^\dagger(t) = e^{i L t} \rightarrow e^{i L_\infty t} = \Pi_\infty^{-1}(t).
\end{align}

\begin{figure*}[t]
	\subfloat{
		\centering
		\includegraphics[width=0.45\linewidth]{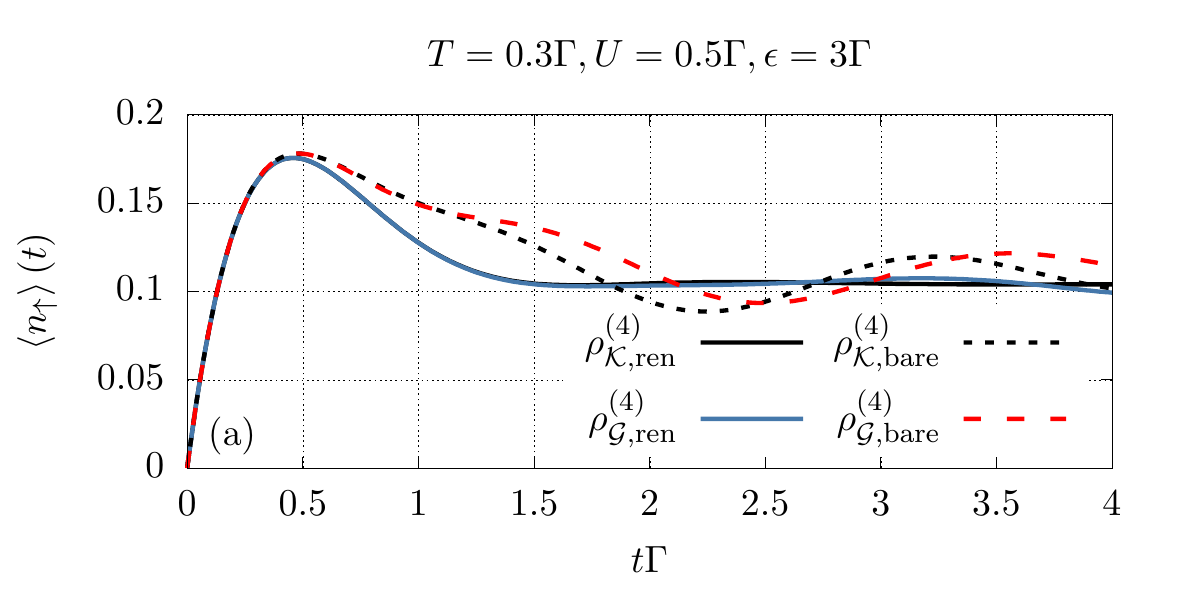}}
	\subfloat{
		\centering
		\includegraphics[width=0.45\linewidth]{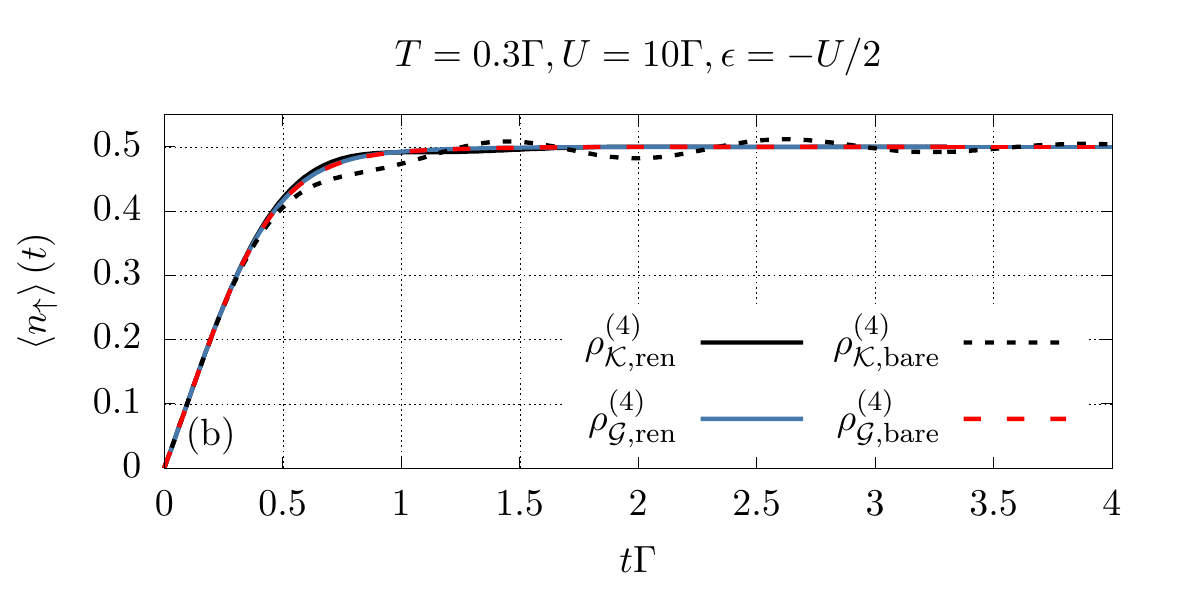}} \\
	\subfloat{
		\centering
		\includegraphics[width=0.45\linewidth]{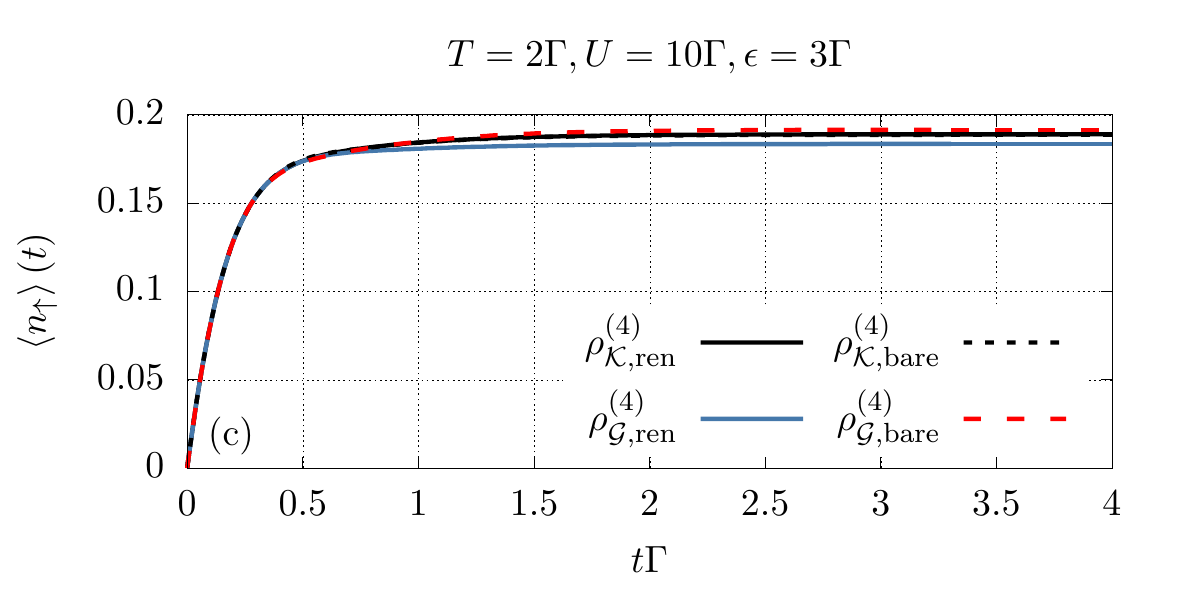}
	}
		\subfloat{
		\centering
		\includegraphics[width=0.45\linewidth]{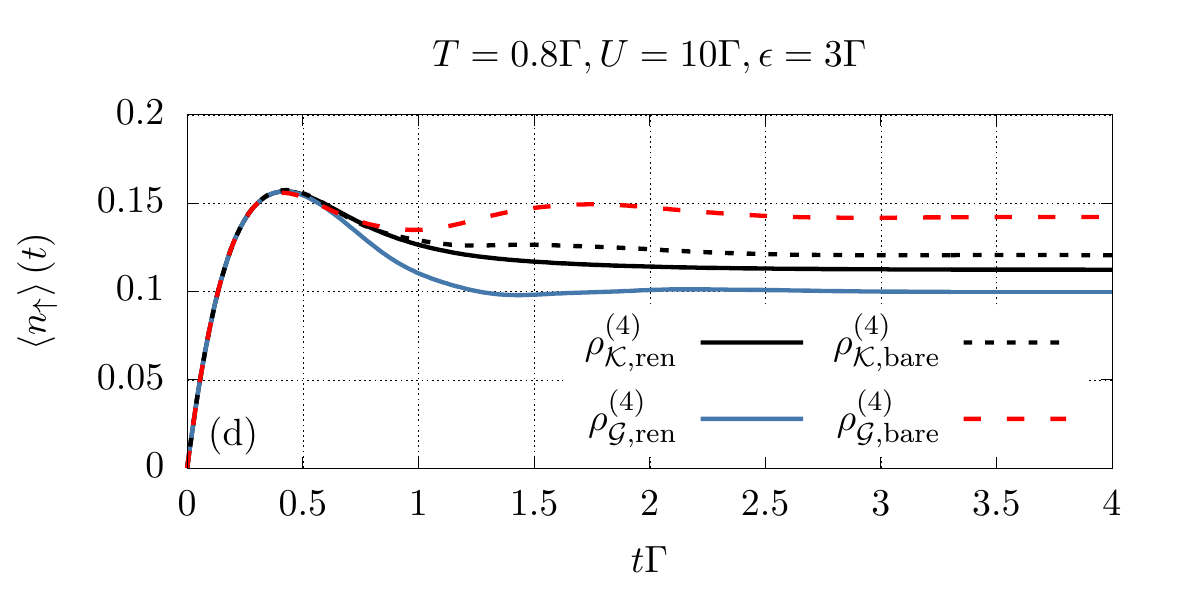}
	} \\
	\subfloat{
		\centering
		\includegraphics[width=0.45\linewidth]{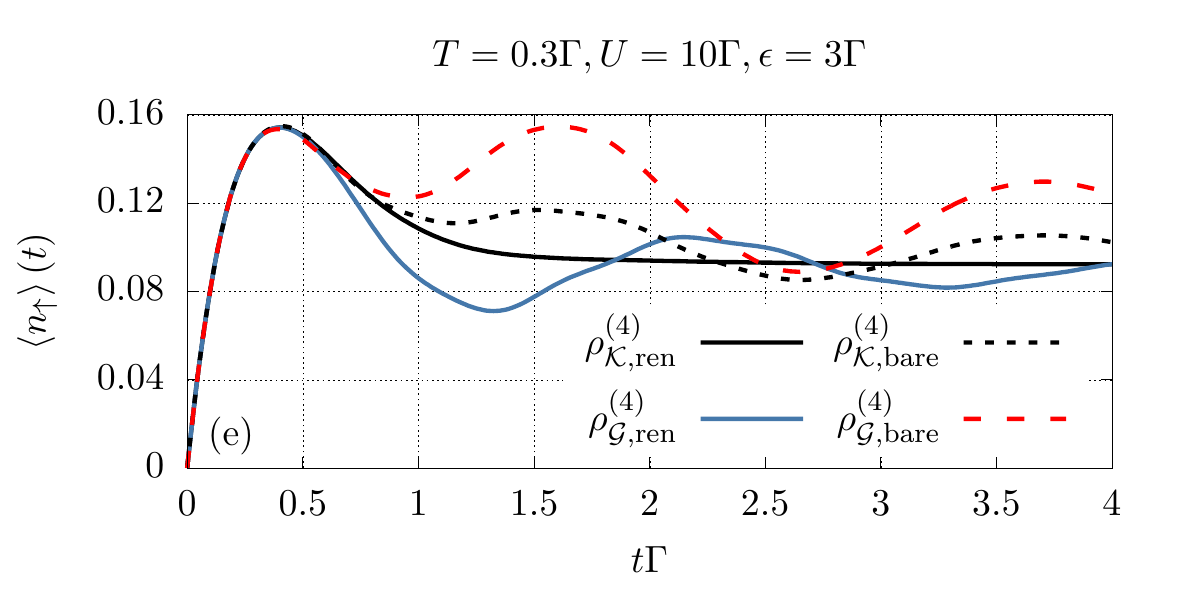}
	}
	\subfloat{
		\centering
		\includegraphics[width=0.45\linewidth]{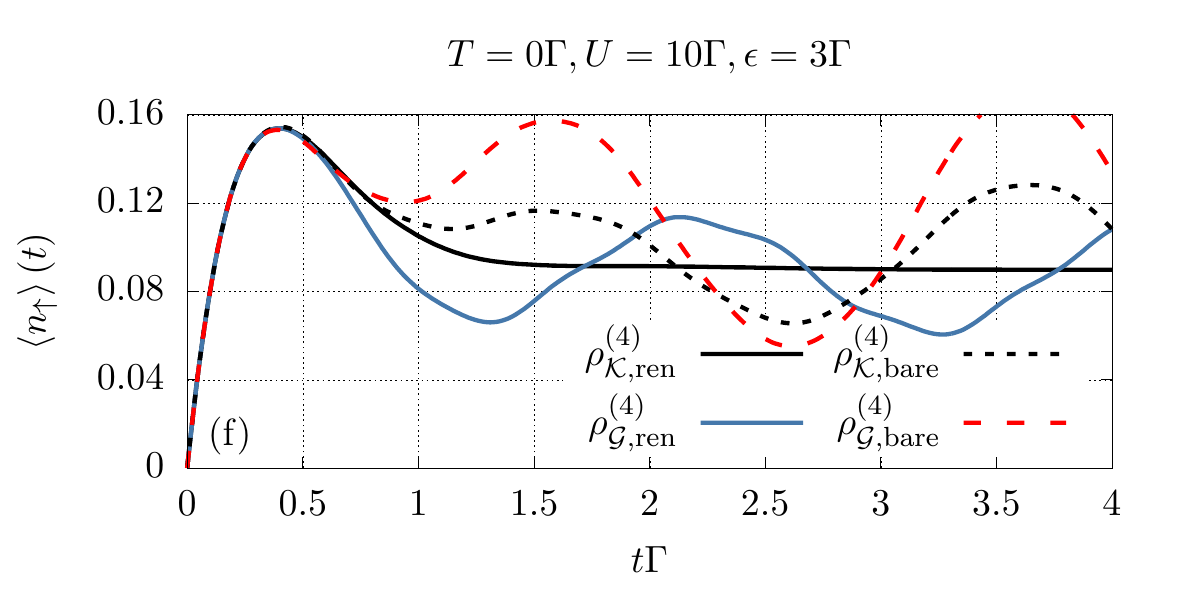}
	}
	\caption{
		Interacting Anderson dot with $\mu_L = \mu_R = 0$: level occupation ($\langle n_\uparrow \rangle =\langle n_\downarrow \rangle$) from the initially unoccupied state.\\ (a) Weak interaction, low temperature. (b) Strong interaction and low temperature at the symmetry point. (c)-(f) Strong interaction with decreasing temperatures off-resonance.
	}
	\label{fig:bare_and_ren_nz}
\end{figure*}

For the $U=0$ limit this implies that because 
the renormalized series for $\K$ terminates at fourth order to give the exact result,
the fourth order generator $\G=L_\infty + \G^{(2)}_\text{ren} + \G^{(4)}_\text{ren}$ is also exact at $U=0$. Surprisingly however, it turns out that at $U=0$ the fourth order contribution \emph{also} vanishes, $\G^{(4)}_\text{ren}=0$, which we verify in \App{app:G2_exact}. Thus, we conclude that the leading-order renormalized generator $\G=L_\infty + \G^{(2)}_\text{ren}$ is already exact for $U=0$, \emph{one order lower} then for the memory kernel $\K=L_\infty + \K^{(2)}_\text{ren} + \K^{(4)}_\text{ren}$.

In \Fig{fig:bare_and_ren_nz} we compare the different fourth order methods.
In \Fig{fig:bare_and_ren_nz}(a) we show results at low temperature $T<\Gamma$ and small interaction $U<\Gamma$. We see that the renormalized solutions coincide, but clearly differ from the bare solutions, which are distinct. Initially the bare solutions also coincide (up to $t\Gamma \simeq 1$) and decay like the renormalized ones, but at a smaller rate. They then start to oscillate while their renormalized counterparts have already reached their stationary values. The stationary values are similar for all methods.
We focused on the off resonant case $\epsilon > \Gamma > U \simeq T$ since here the fourth order corrections are important, also in the renormalized methods.

In \Fig{fig:bare_and_ren_nz}(b) we verify that at the symmetry point $\epsilon = -U/2$ the occupations converge to the same stationary value $n_{\uparrow}=n_{\downarrow}=\tfrac{1}{2}$ as they should by symmetry. However, the bare $\K$ solution predicts an oscillation, which is not predicted by the other methods. By contrast, the bare $\G$ method agrees with the renormalized methods showing no oscillations, even though $U=10\Gamma$ is rather large.

It is thus interesting to consider less constrained parameters with detuning $\epsilon \gtrsim \Gamma$ in the tail of the resonance.
For $U>\Gamma$ we find that at high temperature, $T>5\Gamma$, all methods coincide for these parameters. However, in \Fig{fig:bare_and_ren_nz}(c) we see that already for $T=2\Gamma$ the renormalized $\G$ method predicts a \emph{different} stationary occupation.

When lowering the temperature further in \Fig{fig:bare_and_ren_nz}(d) we see that all methods after the initial rise predict decay ($t\Gamma\gtrsim\tfrac{1}{2}$), except that the renormalized $\G$ method gives a larger rate. Whereas the renormalized $\K$ method reaches stationarity after this, all other methods show similar overdamped oscillations. Because for the renormalized $\G$ method the initial decay is slower and last longer, this oscillation is out of phase with the other ones. Furthermore, it can be seen that every method predicts a different stationary value.

This picture persists when temperature is lowered further in \Fig{fig:bare_and_ren_nz}(e)-(f). The above mentioned damped oscillations grow, whereas the renormalized $\K$ method further reduces the stationary value without introducing oscillations. The renormalized $\G$ method instead features pronounced oscillations with a visible second harmonic and an approximately $\tfrac{\pi}{2}$ phase shift. At $T=0$ the oscillations of the bare solutions even show negative damping, diverging at long times. This means that one of the complex frequency poles of the propagator has moved into the unphysical part of the complex plane~\cite{Schoeller18} and the stationary state is never reached.

The fact that the oscillations in \Fig{fig:bare_and_ren_nz}(e)--(f) don't quickly decay for the renormalized $\G$ method may seem surprising at first,
since the renormalized propagator $\Pi_\infty$ [\Eq{eq:infinite_T_Pi}] overdamps oscillatory contributions to the dynamics as mentioned earlier. However, this is counteracted by insisting on a \emph{time-local} formulation of the renormalized approach. This is immidiately clear from our central result \Eq{eq:G1_from_K}--\eq{eq:total_G_from_K}: in a renormalized expansion with decaying free reference evolution the partial backward time integration will partially undo this decay.

Another way of seeing that time locality is the problem here, is by considering the formal definition of the generator $\G=i \dot\Pi \Pi^{-1}$. In order to obtain any perturbative series for $\G$ it is necessary to expand the \emph{inverse} propagator $\Pi^{-1}$.
For the renormalized expansion \Eqs{eq:G1_from_K}-\eq{eq:total_G_from_K} this involves expanding
\begin{align}
\Pi^{-1} =& \left[ \Pi_\infty + \Pi^{(2)}_{\text{ren}} + \Pi^{(4)}_{\text{ren}} + \cdots \right]^{-1} \\
            =& \, \Pi^{-1}_\infty - \Pi^{-1}_\infty \Pi^{(2)}_{\text{ren}} \Pi^{-1}_\infty + \cdots.
\end{align}
Since $\Pi_\infty$ contains oscillating and decaying contributions (from $L$ and $\Sigma_{\infty}$ respectively [\Eq{eq:infinite_T_Pi}]), it follows that $\Pi_\infty^{-1}$ is exponentially increasing in time. However, the geometric series is only guaranteed to converge if 
\begin{align}
\Big\lVert \Pi^{-1}_\infty \left( \Pi^{(2)}_{\text{ren}} + \Pi^{(4)}_{\text{ren}} + \cdots  \right) \Big\rVert < 1.
\label{eq:convergence_condition}
\end{align}
Because $\Pi^{(2)}_{\text{ren}}(t) + \Pi^{(4)}_{\text{ren}}(t) + \cdots$ converges to a stationary (non-zero) value $\Pi^{(2)}_{\text{ren}}(\infty) + \Pi^{(4)}_{\text{ren}}(\infty) + \cdots$ condition \eq{eq:convergence_condition} is violated after a short time and the time local generator becomes problematic. Note carefully that only for $U=0$ no problems arise with the renormalized $\G$, because higher order corrections are identically zero by the algebraic structure of the model [\App{app:G2_exact}] and convergence of $\G$ is not an issue.
By contrast, in the bare perturbation theory for $\G$ discussed in \Sec{sec:bare_G} this problem did not occur, because the unitary reference evolution $\big\lVert \Pi_0^\dagger \big\rVert$ is always bounded. This shows that the application of renormalized perturbation expansions is much more subtle in the \emph{time-local} framework than in the time-nonlocal one. 
This seems to be a generic problem of any perturbative expansion of $\G$ around a reference solution that already incorporates some dissipative/decaying behavior. This is, however, a key idea behind renormalization strategies for open system~\cite{Schoeller09, Schoeller18, Pletyukhov12a, Lindner18, Lindner19}, which through their use of the kernel $\K$ suffer no such failure. It remains an intriguing open questions whether similar schemes can be developed for $\G$.

 \section{Summary\label{sec:summary}}

We have shown that the recently discovered~\cite{Nestmann21a} \emph{fixed-point} relation \eq{eq:fixed_point} between the memory kernel $\K$ and the generator $\G$ implies a \emph{recursive} relation between time-local and time-nonlocal perturbation series based on the \emph{common} expansion reference $\K_L$. This relation can be exploited to set up calculations of these quantities irrespective of the chosen technique (diagrammatics, projection operators, etc.).
Importantly, it allows for an unbiased comparison of the \emph{different} approximations that result when performing the \emph{same} expansion in a time-local or time-nonlocal picture, independent of model specifics.
The flexibility in the choice of expansion reference $\K_L$ allows to compare bare expansions
with renormalized ones.

For the bare expansion
($\K_L
=L=[H,\bullet]$)
discussed in \Sec{sec:bare_perturbation},
we developed a diagrammatic technique for computing the time-local generator $\G$, in close analogy to the well-developed technique for the memory kernel $\K$.
Judging by the very basic criterion of legitimacy of the approximate propagator (complete positivity),
performing the expansion in the time-local formulation leads to a better behaved solution in application to strongly interacting open systems than performing the corresponding expansion in the time-nonlocal one. Combined with its inherent advantages in addressing questions related to
non-Markovianity~\cite{Rivas10,Chruscinski12a,Rivas14,Wissmann15,Bae2016}
and quantum information, this suggests that the time-local approach made more accessible here \emph{via} the standard time-nonlocal one can be a useful alternative to the existing time-local methods~\cite{Tokuyama75,Tokuyama76,Shibata77,Shibata80,Chaturvedi79,Breuer01,BreuerPetruccione}.
We also note that for the time evolution of \emph{transport observables} -- measured outside the system -- similar memory kernels can be calculated using the same standard techniques~\cite{Schoeller09, Schoeller18}. The present paper also provides a starting point for transposing these techniques to the time-local calculation of transport observables.

For the renormalized expansion ($\K_L=L+\Sigma_{\infty}$) that we additionally developed in \Sec{sec:renormalized_perturbation} this advantage of $\G$ over $\K$ at first seems to be confirmed.
Expanding about the infinite temperature limit,
we found that in the time-local framework the non-interacting Anderson dot is \emph{exactly} solved by the leading order
result, \emph{one order lower} than  in the time-nonlocal framework.
However, in the presence of interaction the unbounded growth of the \emph{dissipative backward} evolution with time leads to problems. We noted that the expansion of the inverse propagator, implicitly required by the expansion of the timelocal generator is questionable on times of the order of the inverse decay rate $\Gamma^{-1}$.

As explained at the end of \Sec{sec:functional}, the versatile fixed-point equation of \Ref{Nestmann21a} may provide an additional route to a renormalized time-local approach: Noting that the renormalized time-nonlocal approach allows to obtain an approximate $\K_\text{pert}$,
iteration of the fixed-point functional \eq{eq:functional_def} can be used to obtain an \emph{equivalent} time-local generator $\G_{\text{sc}}$, which \emph{self-consistently} accounts for the backward evolutions.
This provides an approximation different from the truncated renormalized approach to $\G$ explored here, which follows the traditional approach of expanding $\G$ itself.
Overall, we thus illustrated how the fixed-point relation can be used to transpose standard memory kernel techniques to the interesting but more challenging time-local framework.
Our systematic comparison of the time-local and time-nonlocal framework highlighted their complementary merits and limitations, underscoring the importance of improving our understanding of the connection between these canonical approaches to open-system dynamics.

\acknowledgments
We thank V. Bruch, J. Schulenborg and B. Vacchini for useful discussions.
K.N. acknowledges support by the Deutsche Forschungsgemeinschaft (RTG 1995).

\appendix
\section{Well-definedness of the perturbation theory in second and fourth order\label{app:everything_finite}}

It is at first unclear whether the diagrams in \Eqs{eq:diagram_K_2}--\eq{eq:diagram_K_4_2} and \Eqs{eq:diagram_K_2_ren}--\eq{eq:diagram_K_4_2_ren} of the bare and renormalized $\K$ perturbation theory respectively are actually well defined
because of the singularity in the contraction function $\gamma^{-}_{\eta \sigma}(t)$ [\Eq{eq:contractions}] at $t=0$.
This arises because we have taken the wideband limit from start.
In \Ref{Saptsov12} the bandwidth dependence was discussed [Eq. (75)--(76) loc. cit.] in the frequency representation
but not in the time-representation used here.
Here we specifically show the finiteness of the renormalized perturbation theory up to fourth order using corresponding arguments.
By replacing $L_\infty \rightarrow L$ everywhere the exact same steps establish the finiteness of the bare perturbation theory. First note that $\gamma^{-}_{\eta \sigma}(t)$ diverges as $1/t$ for $t\rightarrow 0$, in particular
\begin{align}
\lim_{t \rightarrow 0} t \, \gamma^{-}_{\eta \sigma}(t) = -i \sum_r \frac{\Gamma_{r \sigma}}{\pi}.
\label{eq:finite_combination_1}
\end{align}
However, because the superfermion superoperators anticommute [\Eq{eq:superfermions_anticommute}] we have
\begin{align}
&\sum_{\eta} \gamma^{-}_{\eta \sigma}(t) G^+_{\eta \sigma} G^+_{\bar\eta \sigma} \label{eq:finite_combination_2} \\
    &\quad= \frac{1}{2} \sum_{\eta} \left[ \gamma^{-}_{\eta \sigma}(t) G^+_{\eta \sigma} G^+_{\bar\eta \sigma} + \gamma^{-}_{\bar\eta \sigma}(t) G^+_{\bar\eta \sigma} G^+_{\eta\sigma} \right] \\
    &\quad= \frac{1}{2} \sum_{\eta} \left[ \gamma^{-}_{\eta \sigma}(t) G^+_{\eta \sigma} G^+_{\bar\eta \sigma} - \gamma^{-}_{\bar\eta \sigma}(t) G^+_{\eta\sigma} G^+_{\bar\eta \sigma} \right] \\
    &\quad= \frac{1}{2} \sum_{\eta} \left[ \gamma^{-}_{\eta \sigma}(t) - \gamma^{-}_{\bar\eta \sigma}(t) \right] G^+_{\eta \sigma} G^+_{\bar\eta \sigma} \\
    &\quad= \frac{-i}{2} \sum_{\eta r} \left[ e^{i \bar\eta \mu_r t} - e^{i \eta \mu_r t}  \right] \frac{\Gamma_{r \sigma} T_r}{\sinh(\pi T_r t)} G^+_{\eta \sigma} G^+_{\bar\eta \sigma} \\
    &\quad= \sum_{\eta r} \sin(\bar\eta \mu_r t) \frac{\Gamma_{r \sigma} T_r}{\sinh(\pi T_r t)} G^+_{\eta \sigma} G^+_{\bar\eta \sigma} \\
    &\quad= -2\sum_{r} \Gamma_{r \sigma} T_r \frac{\sin(\mu_r t)}{\sinh(\pi T_r t)} G^+_{+ \sigma} G^+_{- \sigma}.
\end{align}
Thus, we see that the apparent singularity in \Eq{eq:finite_combination_2} at $t=0$ never contributes. The $t\rightarrow 0$ limit of \Eq{eq:finite_combination_2} is specifically given by
\begin{align}
\lim_{t\rightarrow 0} \sum_{\eta} \gamma^{-}_{\eta \sigma}(t) G^+_{\eta \sigma} G^+_{\bar\eta \sigma} = -2 \sum_r \frac{\Gamma_{r \sigma}}{\pi} \mu_r G^+_{+ \sigma} G^+_{- \sigma}.
\label{eq:finite_combination_3}
\end{align}
We can now rewrite the second order renormalized $\K$ diagram as
\begin{align}
\DiagramKTwo =& -\sum_{\eta \sigma}\gamma^-_{\eta \sigma}(t) G^+_{\eta \sigma} e^{-i L_\infty t} G^{+}_{\bar\eta \sigma} \label{eq:diagram_K_two_again} \\
    =& -\sum_{\eta \sigma}\gamma^-_{\eta \sigma}(t) G^+_{\eta \sigma} \left[e^{-i L_\infty t} - \ones \right] G^{+}_{\bar\eta \sigma} \notag \\
    &-\sum_{\eta \sigma} \gamma^-_{\eta \sigma}(t) G^+_{\eta \sigma} G^{+}_{\bar\eta \sigma}
\end{align}
Using \Eq{eq:finite_combination_1} and \Eq{eq:finite_combination_3} this immidiately shows that this diagram is finite for $t\rightarrow 0$:
\begin{align}
\lim_{t\rightarrow 0}\, \DiagramKTwo =& \sum_{\eta \sigma r} \frac{\Gamma_{r \sigma}}{\pi} G^+_{\eta \sigma} L_\infty G^{+}_{\bar\eta \sigma} + 2 \sum_{\sigma r} \frac{\Gamma_{r \sigma}}{\pi} \mu_r G^+_{+ \sigma} G^+_{- \sigma}.
\label{eq:diagram_K_two_t_zero_limite}
\end{align}

Since the second order diagram is contained within one of the fourth order diagrams [\Eq{eq:diagram_K_4_2_ren}], it follows that
\begin{align}
\DiagramKFourTwo = \mathcal{O}(t) \text{ as } t \rightarrow 0.
\end{align}
This is because the outer contraction $\gamma^{-}_{\eta_1 \sigma_1}(t)$ diverges as $1/t$, but the inner integrals $\int_{0}^t dt_1 \int_{0}^{dt_1} dt_2 \dots = \mathcal{O}(t^2)$ vanish quadratically.
To see the well-definedness of the other fourth order diagram we decompose it as
\begin{align}
&\DiagramKFourOne = F_1(t) + F_2(t), \\
&F_1(t) \coloneqq - \sum_{\eta_1 \sigma_1} \sum_{\eta_2 \sigma_2} \int_0^t dt_1 \int_0^{t_1} dt_2 \gamma^{-}_{\eta_1 \sigma_1}(t-t_2)  \gamma^{-}_{\eta_2 \sigma_2}(t_1) \notag \\
&\quad \times G^+_{\eta_1 \sigma_1} e^{-i L_\infty (t-t_1)} G^+_{\eta_2 \sigma_2} \notag\\
&\quad  \times \left[ e^{-i L_\infty (t_1-t_2)} - \ones \right] G^{+}_{\bar\eta_1 \sigma_1} e^{-i L_\infty t_2} G^{+}_{\bar\eta_2 \sigma_2}, \\
&F_2(t) \coloneqq - \sum_{\eta_1 \sigma_1} \sum_{\eta_2 \sigma_2} \int_0^t dt_1 \int_0^{t_1} dt_2 \gamma^{-}_{\eta_1 \sigma_1}(t-t_2)  \gamma^{-}_{\eta_2 \sigma_2}(t_1) \notag \\
&\quad \times G^+_{\eta_1 \sigma_1} e^{-i L_\infty (t-t_1)} G^+_{\eta_2 \sigma_2} G^{+}_{\bar\eta_1 \sigma_1} e^{-i L_\infty t_2} G^{+}_{\bar\eta_2 \sigma_2}.
\end{align}
The first contraction $\gamma^{-}_{\eta_1 \sigma_1}(t-t_2)$ in $F_1(t)$ diverges if $t_2 \rightarrow t$. But in this limit the factor $e^{-i L_\infty (t_1-t_2)} - \ones$ vanishes with $\mathcal{O}(t_1-t_2)$ because of the time ordering $t\geq t_1 \geq t_2$. Therefore the divergence in $\gamma^{-}_{\eta_1 \sigma_1}(t-t_2)$ is always regularized. The second contraction $\gamma^{-}_{\eta_2 \sigma_2}(t_1)$ in $F_1(t)$ diverges for $t_1 \rightarrow 0$, which is regularized by the inner integral $\int_0^{t_1} dt_2 = \mathcal{O}(t_1)$. Hence $F_1(t)$ is always finite. In the second term $F_2(t)$ one first uses the anticommutation $G^+_{\eta_2 \sigma_2} G^{+}_{\bar\eta_1 \sigma_1}=-G^{+}_{\bar\eta_1 \sigma_1}G^+_{\eta_2 \sigma_2}$. Note again that in the limit $t_2 \rightarrow t$ we also have $t_1 \rightarrow t$ because of the time ordering. This means that the factor
\begin{align}
\gamma^{-}_{\eta_1 \sigma_1}(t-t_2) G^+_{\eta_1 \sigma_1} e^{-i L_\infty (t-t_1)} G^{+}_{\bar\eta_1 \sigma_1}
\end{align}
is always finite for $t_2 \rightarrow t$ following the same argument which established that \Eq{eq:diagram_K_two_again} has the finite limit \eq{eq:diagram_K_two_t_zero_limite}. For precisely the same reason the other factor
\begin{align}
\gamma^{-}_{\eta_2 \sigma_2}(t_1) G^+_{\eta_2 \sigma_2} e^{-i L_\infty t_2} G^{+}_{\bar\eta_2 \sigma_2}
\end{align}
is finite for $t_1 \rightarrow 0$. Therefore $F_2(t)$ is also always finite, establishing the well-definedness of this last diagram. \section{Exact generator at $U=0$\label{app:G2_exact}}

Here we show that the renormalized \emph{second} order generator $\G = L_\infty + \G^{(2)}_{\text{ren}}$ is already exact in the noninteracting case by showing that the fourth order correction is identically zero, $\G^{(4)}_{\text{ren}} = 0$, by a nontrivial cancellation of terms. To do so we split up the fourth order renormalized generator into two contributions
\begin{align}
-i \G^{(4)}_{\text{ren}}(t) =& A_1(t) - A_2(t), \\
A_1(t) \coloneqq& \DiagramGFourTwo \cdot \Pi_\infty^{-1} \\
&+ \DiagramGFourOne \cdot \Pi_\infty^{-1},\\
A_2(t) \coloneqq& \DiagramGTwo \cdot \Pi_\infty^{-1} \cdot \DiagramPiTwo \cdot \Pi_\infty^{-1} \\
&- \DiagramGFourThree \cdot \Pi_\infty^{-1}.
\end{align}
and show that $A_1(t) = A_2(t)$.
The main simplification for $U=0$ is that the renormalized free Liouvillian $L_\infty$ and the superfermions satisfy the commutation relation
\begin{align}
\Big[L_\infty, G^+_{\eta \sigma}\Big] = \Big( \eta \epsilon - \tfrac{i}{2} \sum_r \Gamma_{r \sigma} \Big) G^+_{\eta \sigma},
\end{align}
see Eq.~(118) in \Ref{Saptsov14}. From this it follows that
\begin{align}
G_{\eta \sigma}'(t) \coloneqq& e^{i L_\infty t} G^+_{\eta \sigma} e^{-i L_\infty t} \\
 =& e^{\big(i \eta \epsilon + \tfrac{1}{2} \sum_r \Gamma_{r \sigma}  \big) t} G^+_{\eta \sigma} .
\end{align}
These transformed superfermions still anticommute $\left\{G_{\eta_1 \sigma_1}'(t_1), G_{\eta_2 \sigma_2}'(t_2)\right\} = 0$. 
We now rewrite $A_1(t)$ using the $G'_{\eta \sigma}$ as
\begin{widetext}
\begin{align}
A_1(t) =& e^{-i L_\infty t} \underset{t > t_1 > t_2 > t_3 > 0}{\int dt_1 dt_2 dt_3} 
    \Big[ \gamma_{\eta_1 \sigma_1}(t - t_3) \gamma_{\eta_2 \sigma_2}(t_1 - t_2) G_{\eta_1 \sigma_1}'(t) G_{\eta_2 \sigma_2}'(t_1) G_{\bar\eta_2 \sigma_2}'(t_2) G_{\bar\eta_1 \sigma_1}'(t_3) \Big. \notag \\
    & \qquad\qquad\qquad\qquad- \gamma_{\eta_1 \sigma_1}(t - t_2) \gamma_{\eta_2 \sigma_2}(t_1 - t_3) G_{\eta_1 \sigma_1}'(t) G_{\eta_2 \sigma_2}'(t_1) G_{\bar\eta_1 \sigma_1}'(t_2) G_{\bar\eta_2 \sigma_2}'(t_3)
    \Big. \Big] e^{i L_\infty t} \label{eq:expansion_A_1_1} \\
    =& e^{-i L_\infty t} \underset{t > t_1 > t_2 > t_3 > 0}{\int dt_1 dt_2 dt_3}
    \Big[ \gamma_{\eta_1 \sigma_1}(t - t_3) \gamma_{\eta_2 \sigma_2}(t_1 - t_2) G_{\eta_1 \sigma_1}'(t) G_{\bar\eta_1 \sigma_1}'(t_3) G_{\eta_2 \sigma_2}'(t_1) G_{\bar\eta_2 \sigma_2}'(t_2) \Big. \notag \\
    & \qquad\qquad\qquad\qquad+ \gamma_{\eta_1 \sigma_1}(t - t_2) \gamma_{\eta_2 \sigma_2}(t_1 - t_3) G_{\eta_1 \sigma_1}'(t) G_{\bar\eta_1 \sigma_1}'(t_2) G_{\eta_2 \sigma_2}'(t_1) G_{\bar\eta_2 \sigma_2}'(t_3)
    \Big. \Big] e^{i L_\infty t} \label{eq:expansion_A_1_2} \\
    =& e^{-i L_\infty t} \Big[ \underset{t > t_2 > t_3 > t_1 > 0}{\int dt_2 dt_3 dt_1} + \underset{t > t_2 > t_1 > t_3 > 0}{\int dt_2 dt_1 dt_3} \Big]
    \gamma_{\eta_1 \sigma_1}(t - t_1) \gamma_{\eta_2 \sigma_2}(t_2 - t_3) G_{\eta_1 \sigma_1}'(t) G_{\bar\eta_1 \sigma_1}'(t_1) G_{\eta_2 \sigma_2}'(t_2) G_{\bar\eta_2 \sigma_2}'(t_3)
     e^{i L_\infty t} \label{eq:expansion_A_1_3} \\
     =& e^{-i L_\infty t} \int_0^t dt_1 \int_{t_1}^t dt_2 \int_{0}^{t_2} dt_3
     \gamma_{\eta_1 \sigma_1}(t - t_1) \gamma_{\eta_2 \sigma_2}(t_2 - t_3) G_{\eta_1 \sigma_1}'(t) G_{\bar\eta_1 \sigma_1}'(t_1) G_{\eta_2 \sigma_2}'(t_2) G_{\bar\eta_2 \sigma_2}'(t_3)
     e^{i L_\infty t}
\end{align}
From \Eq{eq:expansion_A_1_1} to \Eq{eq:expansion_A_1_2} we used the anticommutation property of the $G'_{\eta \sigma}$. From \Eq{eq:expansion_A_1_2} to \Eq{eq:expansion_A_1_3} we relabeled the integration variables in the first term as $t_1 \rightarrow t_2 \rightarrow t_3 \rightarrow t_1$ and in the second term as $t_1 \leftrightarrow t_2$. But for $A_2(t)$ we have similarly
\begin{align}
A_2(t) =& e^{-i L_\infty t} \Big[ \int_0^t dt_1 \int_0^t dt_2 \int_0^{t_2} dt_3 - \underset{t > t_1 > t_2 > t_3 > 0}{\int dt_1 dt_2 dt_3} \Big] \gamma_{\eta_1 \sigma_1}(t - t_1) \gamma_{\eta_2 \sigma_2}(t_2 - t_3) G_{\eta_1 \sigma_1}'(t) G_{\bar \eta_1 \sigma_1}'(t_1) G_{\eta_2 \sigma_2}'(t_2) G_{\bar\eta_2 \sigma_2}'(t_3) e^{i L_\infty t}\\
    =& e^{-i L_\infty t} \int_0^t dt_1 \int_{t_1}^{t} dt_2 \int_0^{t_2} dt_3 \gamma_{\eta_1 \sigma_1}(t - t_1) \gamma_{\eta_2 \sigma_2}(t_2 - t_3) G_{\eta_1 \sigma_1}'(t) G_{\bar \eta_1 \sigma_1}'(t_1) G_{\eta_2 \sigma_2}'(t_2) G_{\bar\eta_2 \sigma_2}'(t_3) e^{i L_\infty t}.
\end{align}
\end{widetext}
By comparison we thus see that $A_1(t)=A_2(t)$. Hence $\G^{(4)}_{\text{ren}}=0$. 

 {}

\end{document}